\documentstyle[11pt]{article}

\input epsf
\begin{document}

\newcommand{\beq}{\begin{equation}}
\newcommand{\eeq}{\end{equation}}
\newcommand{\bea}{\begin{eqnarray}}
\newcommand{\eea}{\end{eqnarray}}

\def\bsigma{\mbox{\boldmath $\sigma$}} 
\def\balpha{{\mbox{\boldmath $\alpha$}}}
\def\bbeta{{\mbox{\boldmath $\beta$}}}
\def\bgamma{{\mbox{\boldmath $\gamma$}}}
\def\simle{\mathrel{\mathpalette\fun <}}
\def\simge{\mathrel{\mathpalette\fun >}}
\def\fun#1#2{\lower3.6pt\vbox{\baselineskip0pt\lineskip.9pt
  \ialign{$\mathsurround=0pt#1\hfil##\hfil$\crcr#2\crcr\sim\crcr}}}

\def\Tr{{\rm Tr}\,}
\def\high{\vphantom{\Biggl(}\displaystyle}
\makeatletter
\renewcommand\theequation{\thesection.\arabic{equation}}
\@addtoreset{equation}{section}
\makeatother

\baselineskip 20pt
\rightline{CU-TP-946}
\rightline{hep-th/9908095}
\vskip 2cm

\centerline{{\Large\bf Massive and Massless Monopoles and
Duality}\footnote{Talks delivered at the  Asia Pacific Center for
Theoretical Physics Third Winter School on Duality in Fields and
Strings, Seogwipo, Cheju, Korea, January 1999}}
\bigskip
\centerline{ Erick J. Weinberg}
\medskip
\centerline{ Department of Physics, Columbia University}
\centerline{ New York, NY 10027, USA}
 
\bigskip
\centerline{\bf Abstract}
\medskip
\centerline{\vbox{\hsize=5truein 
\baselineskip=12truept \tenrm \noindent I review some aspects of BPS
magnetic monopoles and of electric-magnetic duality in theories with
arbitrary gauge groups.  When the symmetry is maximally broken to a
U(1)$^r$ subgroup, all magnetically charged configurations can be
understood in terms of $r$ species of massive fundamental monopoles.
When the unbroken group has a non-Abelian factor, some of these
fundamental monopoles become massless and can be viewed as the duals
to the massless gauge bosons.  Rather than appearing as distinct
solitons, these massless monopoles are manifested as clouds of
non-Abelian field surrounding one or more massive monopoles.  I
describe in detail some examples of solutions with such clouds.}}

\baselineskip 20pt

\section{Introduction}

Elementary treatments of quantum field theory usually introduce
particles as the quanta of the small oscillation of a weakly coupled
field about the vacuum.  There is one particle species for each
elementary field, with the masses of the particles entering as
parameters in the Lagrangian.  However, it turns out that particles
can also arise in a very different fashion.  In many cases, the
classical field equations of the theory possess finite energy
solutions that are localized in space.  These solitons also give rise
to one-particle states in the quantum theory.  To lowest
approximation, the mass of the soliton is equal to the energy of the
classical solution and is typically of the form $M_{\rm soliton} \sim
m/\lambda$, where $m$ is an elementary particle mass and $\lambda$ is
some small coupling.

At first sight, these two types of particles seem radically different.
The elementary excitations appear to be point particles with no
substructure and no internal degrees of freedom.  The solitons, on the
other hand, are extended objects characterized by a classical field
profile $\phi({\bf x})$ that is meaningful in the quantum theory
because its spatial extent ($\sim 1/m$) is much greater than the
Compton wavelength ($\sim \lambda/m$) of the soliton.

On closer inspection, however, this distinction is less clearcut.  On
the one hand, in the presence of interactions the ``elementary''
particles turn out to be not entirely structureless; instead, they
have a partonic substructure that evolves with momentum scale.  On the
other hand, the internal structure of the soliton appears somewhat
simpler when it is analyzed in terms of normal modes rather than
directly in terms of the classical solution.  Most of the small
fluctuation modes are nonnormalizable modes, with continuum
eigenvalues, that are most naturally understood as scattering states
of the elementary quanta in the presence of the soliton.  There may
also be normalizable modes with nonzero eigenvalues; if so, these
correspond to states with elementary quanta bound to the soliton.
This leaves only a small number of normalizable zero eigenvalue modes,
whose quantization requires the introduction of collective
coordinates.  It is only these that correspond to fundamental degrees
of freedom of the soliton.

These observations suggest that the one-particle states built from
solitons and those based on the elementary quanta might not differ in
any essential way, and that the apparent differences between the two
may be simply artifacts of the weak coupling regime.  If so, then what
happens as the coupling is increased?  For small coupling, where the
semiclassical treatment of the soliton is still valid, we know that
the ratio of the soliton mass to the elementary particle mass
decreases, as indicated in Fig.~1.  However, such perturbative results
are no longer reliable by the time the couplings have become of order
unity, the region indicated by the question mark.  Nevertheless, 
\vskip 5mm
\begin{center}
\leavevmode
\epsfysize =2in\epsfbox{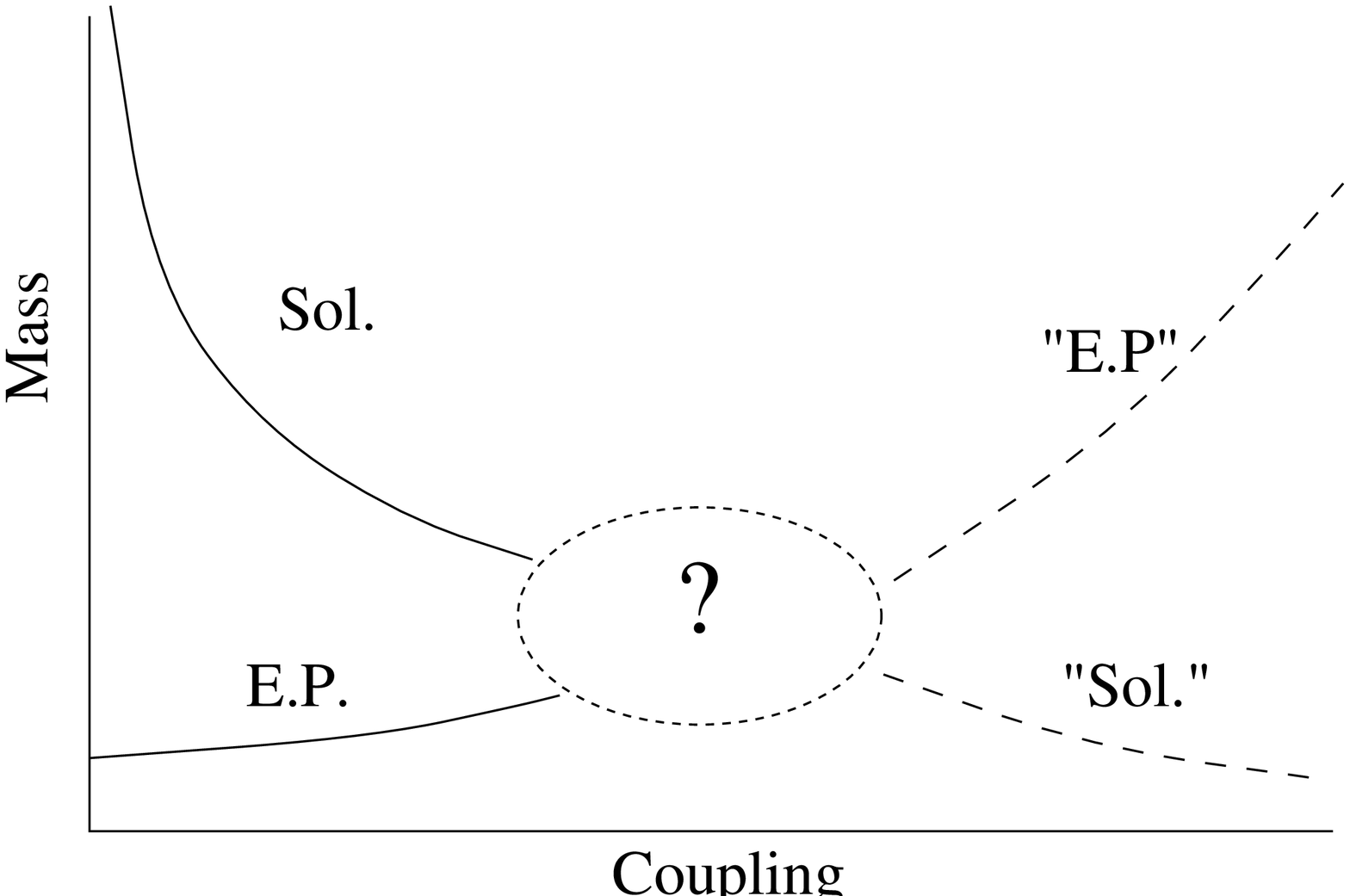}
\end{center}
\begin{quote}
{\bf Figure 1:} {\small  Possible evolution of elementary (E.P.) and
soliton (Sol.) masses with increasing coupling strength.}
\end{quote}
\vskip 0.38cm
\noindent one
can speculate.  For example, it might happen, as indicated in the
figure, that the soliton mass continues to decrease until, at some
large coupling, the ``soliton'' is much lighter than the ``elementary
particle''.  Of course, as suggested by the quotation marks, in this
large coupling regime there is no reason to expect the corresponding
states to have any of the characteristics that they had at small
coupling.

The symmetry between the left and right sides of this figure suggests
a particularly intriguing possibility.  Could the theory be
reformulated so that the roles of the ``soliton'' and the ``elementary
particle'' are interchanged?  In some cases, the answer is positive:
There is a dual formulation of the theory in terms of a new set of
fields whose elementary excitations give rise to the soliton states of
the original formulation, while the former elementary particles
correspond either to solitons or to bound states.  Weak
coupling for this new formulation corresponds to strong coupling for
the original.  This duality can by summarized by
\begin{eqnarray}
   {\rm strong~coupling}~{\cal L}_1(\phi) &\longleftrightarrow & {\rm
       weak~coupling}~{\cal L}_2(\chi) \nonumber\\ 
     {\rm  elementary}~\phi &\longleftrightarrow & {\rm soliton}
       \nonumber\\ 
     {\rm soliton}&\longleftrightarrow & {\rm  elementary}~\chi 
             \nonumber 
\end{eqnarray}

Note that there is no reason for the dual Lagrangian ${\cal L}_2$ to
have the same form as the original Lagrangian ${\cal L}_1$.  For
example, in the classic illustration of field theory duality, that
between the sine-Gordon and the massive Thirring models \cite{sineG},
there seems to be no resemblance at all between the two theories.
However. it could happen that ${\cal L}_1$ and ${\cal L}_2$ have the
same functional form.  As I will describe in more detail below, it is
believed that certain supersymmetric Yang-Mills theories have a
self-duality of this kind that interchanges electric and
magnetic charges and that can be seen as a generalization of the
duality symmetry of Maxwell's equations \cite{dual}.

Whether or not there is a dual formulation of the theory, these
correspondences between the elementary particle and the soliton states
of the quantum theory have implications for the structure and
properties of the classical soliton solutions.  Quite often, these
solutions are associated with a conserved topological charge.  In many
such cases, one finds not only a solution with unit topological
charge, but also families of multiply-charged classical solutions.  If
the soliton one-particle states of the quantum theory are not
fundamentally different from those built from the elementary
excitations, then one might expect that the multiply-charged solutions
would correspond to multiparticle states.  This expectation is clearly
borne out in a number of examples, including, as I will describe in
these lectures, the magnetic monopole solutions in many spontaneously
broken gauge theories.  However, we will also see that in some cases
--- those where the unbroken gauge group has a non-Abelian component
--- where this correspondence is less clearcut.  In these theories the
elementary particle sector includes massless particles carrying
electric-type charges.  The duals of these should be massless
magnetically charged objects.  Such massless monopoles do not appear
as isolated classical solutions.  Instead, they lose their individual
identity and are manifested as ``clouds'' of non-Abelian fields that
surround the massive monopoles \cite{nonabelian}.  Understanding the
nature of these clouds may well provide insight into the properties of
non-Abelian gauge theories.

The remainder of these lectures is organized as follows.  In Sec.~2, I
describe the classic example of field theory duality, that between the
sine-Gordon and massive Thirring models in two spacetime dimensions.
I then move on to four dimensions and the conjectured duality between
magnetic monopoles and electrically charged fundamental excitations.
I begin in Sec.~3 by reviewing the essential properties of the `t
Hooft-Polyakov \cite{hooft} monopole of SU(2) gauge theory.  In
Sec.~4, I describe the Bogomolny-Prasad-Sommerfield \cite{bps}, or
BPS, limit and its relevance for electric-magnetic duality.  Next, in
Sec.~5, I discuss the solutions with higher magnetic charge in an
SU(2) theory and show how these can be understood as corresponding to
multiparticle states.  Section~6 describes how these methods can be
extended to theories with larger gauge groups, concentrating initially
on cases where the symmetry is maximally broken to an Abelian
subgroup.  A very important tool for carrying this out is the moduli
space approximation \cite{manton}, described in Sec.~7.  In Sec.~8, I
describe the special issues that arise when the symmetry breaking is
nonmaximal and the unbroken group is non-Abelian.  It is here that one
finds the massless monopoles clouds.  I describe in detail two
examples of this, a relatively simple one in an SO(5) theory
\cite{nonabelian,so5} in Sec.~9, and a more complex family of
solutions \cite{nahmpaper} in Sec.~10.  Section~11 contains some
concluding remarks and discussion.

\section{The Sine-Gordon -- Massive Thirring Model Duality}

The classic example of duality between two quantum field theories is
the famous equivalence \cite{sineG} between the sine-Gordon and massive
Thirring models.  The sine-Gordon model is a theory of a single scalar
field in $(1+1)$-dimensional spacetime, with Lagrangian density
\begin{equation}
   {\cal L} = {1\over 2} \left(\partial_\mu \phi\right)^2 + {m^2\over
   \beta^2}    \cos \beta\phi \,\, .
\end{equation}
There are an infinite number of degenerate vacua corresponding to the
minima of the scalar field potential at $\phi = 2\pi n/\beta$.
Expanding about one of these (say $\phi=0$), we obtain 
\begin{equation}
   {\cal L} = {1\over 2} \left(\partial_\mu \phi\right)^2 + {m^2\over
 \beta^2}  - {m^2 \over 2} \phi^2 + {m^2 \beta^2 \over 24} \phi^4 + \cdots
\end{equation}
from which we see that $\beta$ is a dimensionless coupling constant,
while $m$ is the lowest-order approximation to the mass of the 
elementary $\phi$ boson.

The existence of discrete degenerate vacua gives rise to a conserved
topological charge 
\begin{equation}
    Q_{\rm top} = {\beta \over 2 \pi} \left[\phi(\infty)-\phi(-\infty)
    \right] 
\end{equation}
that takes on integer values.  There is a classical static soliton
solution --- the ``kink'' --- that approaches two adjacent vacua as $x
\rightarrow \pm\infty$, and thus carries $Q_{\rm top}=1$; there is
also an antikink, with $Q_{\rm top}=-1$.  The corresponding
one-particle states of the quantum theory have masses
\begin{equation}
    M_{\rm kink} = {8 m \over \beta^2} \left(1 - {\beta^2 \over
                        8\pi}\right) \,\, .
\end{equation}
(The prefactor is the classical energy, while the quantity in brackets
is the result of quantum corrections\footnote{I am neglecting here
technical points associated with the renormalization of parameters and
the definitions of composite operators.  For a more rigorous treatment
of these, see \cite{sineG}.}.)

In addition, there are periodic time-dependent classical solutions,
known as breathers.  These resemble a kink and an antikink
bound together and oscillating back and forth.  Using WKB methods,
Dashen, Hasslacher and Neveu \cite{dhn} showed that in the quantum
theory these breathers 
give rise to a series of states with masses
\begin{equation}
    M_n = 2  M_{\rm kink}\,  \sin\left[{n\pi \over 2} \, {\beta^2/8\pi
    \over   1 - \beta^2/8\pi} \right]            
\end{equation}
where $n$ is a positive integer obeying 
\begin{equation}
     n < {8 \pi \over \beta^2} -1 \,\, .
\label{nbreathermax} 
\end{equation}

At this point, it might seem as if there are three classes of states
--- the elementary $\phi$, the kink and antikink, and the breather
states.  Actually, the elementary $\phi$ is the same as the lowest
breather state, as is suggested by the small-$\beta$ expansion
\begin{equation}
   M_n = n m \left[ 1 - {n^2 \beta^4 \over 1536} + O(\beta^6) \right]
\,\, .
\end{equation}

Thus, for small $\beta$ the particle spectrum includes an elementary
boson with mass $M_\phi=M_1$ and $Q_{\rm top}=0$, two solitons with
mass $M_{\rm kink} \gg M_\phi$ and $Q_{\rm top}=\pm 1$, and a series
of $Q_{\rm top}=0$ states with masses $M_\phi < M_n < 2M_{\rm kink}$.
As $\beta$ is increased, the soliton mass decreases, while those of
the elementary $\phi$ and of the higher breather states all increase.
The breather states disappear in succession as their masses 
become greater that twice the soliton mass. Finally, the
elementary $\phi$ disappears when $\beta=\sqrt{4\pi}$, leaving only
the $Q_{\rm top}=\pm 1$ states. 

It is instructive to examine the behavior of the $\phi$ mass as $\beta
\rightarrow \sqrt{4\pi}$.  If we write
\begin{equation}
    {\beta^2\over 4 \pi} = {1 \over 1 + \delta/\pi}    
\end{equation}
the $\phi$ mass is 
\begin{equation}
    M_1 = M_{\rm kink} \left[2 -\delta^2 + {4 \delta^3 \over \pi} 
         + O(\delta^4)\right] \,\, .
\label{SGboundmass}
\end{equation}
This formula suggests that in this large-$\beta$ regime it might be
more natural to regard the kink and antikink --- which are now the
lightest particles in the theory --- as elementary objects, and the
$\phi$ as being a kink and an antikink bound together by a weak
interaction with a strength characterized by $\delta$.  (Note that in
this regime the terms ``kink'' and ``antikink'' are somewhat
misleading, since the semiclassical correspondence between field
profile and quantum state is no longer valid at such strong coupling.)

Let us turn now to the massive Thirring model.  This is a theory of a
fermion field $\psi$, also in $(1+1)$-dimensional spacetime, with
Lagrangian density
\begin{equation}
     {\cal L} =  \bar\psi ( i\gamma^\mu\partial_\mu -M ) \psi 
      - {g\over 2} (\bar\psi \gamma^\mu \psi)^2\,\, .
\end{equation}
The particle spectrum includes an elementary fermion
and its antiparticle, each with mass $M$.  If $g$ is positive, so that
the four-fermion interaction is attractive, there are also
fermion-antifermion bound states.  The lowest (and, for sufficiently
weak coupling, the only) one of these has a mass
\begin{equation}
    M_{\rm Bound} = M \left[2 - g^2 + {4 g^3 \over \pi} + O(g^4)
    \right] \,\, .
\label{MTboundmass}
\end{equation}

The similarity between Eqs.~(\ref{SGboundmass}) and
(\ref{MTboundmass}) suggests that the massive Thirring model might
indeed be a dual theory in which the interpretation of the $\phi$ as a
kink-antikink bound state is realized.  Let us therefore identify $g$
and $\delta$, so that
\begin{equation}
     {\beta^2 \over 4\pi} = {1 \over 1 + g/\pi}  \,\, .
\end{equation}
Weak coupling ($\beta \rightarrow 0$) for the sine-Gordon theory
thus corresponds to strong coupling ($g \rightarrow \infty$) for the
massive Thirring model, while $g\rightarrow 0$ in the latter
corresponds to the strong coupling limit $\beta \rightarrow
\sqrt{4\pi}$ of the former.  We would then have the equivalences
\begin{eqnarray}
   {\rm kink} &\longleftrightarrow & {\rm elementary~}\psi
            \nonumber\\
   {\rm antikink} &\longleftrightarrow & {\rm elementary~}\bar\psi
            \nonumber\\
   {\rm elementary~}\phi&\longleftrightarrow & \bar\psi{\rm -}\psi
     {\rm~bound~state}  \nonumber
\end{eqnarray}
between the particle states.  These, in turn, imply the identification
$$
   {\rm topological~charge} \longleftrightarrow  {\rm
   fermion~number}   
$$
between the conserved charges.

The remarkable fact is that one can beyond these correspondences
between the mass spectra and rigorously establish an equivalence
between the two theories.  Making the operator identifications~\cite{sineG}
\begin{eqnarray} 
       {\beta \over 2\pi} \epsilon^{\mu\nu} \partial_\nu \phi &=&
                - \bar \psi \gamma^\mu \psi \nonumber\\
       {m^2 \over \beta^2} \cos \beta\phi &=& -M \bar \psi\psi
                \nonumber\\
\end{eqnarray}
(with appropriate renormalization and normal ordering of the composite
operators), one finds that the matrix elements of the two theories
agree to all orders of perturbation theory.  Furthermore, the fermion
field operator $\psi({\bf x})$ of the massive Thirring model can be
written \cite{mandelstam} as a nonlocal function of the sine-Gordon boson field
$\phi({\bf x})$.

\section{Magnetic monopoles}
\label{magmon}

Let us now go on to four spacetime dimensions, where the most
important examples of solitons arise in spontaneously broken
gauge theories.  If the vacuum expectation value of the Higgs field
breaks a gauge group $G$ down to a subgroup $H$, there are
topologically stable solitons if the second homotopy group
$\Pi_2(G/H)$ is nonzero.  These are generally referred to as magnetic
monopoles although, strictly speaking, the term is only appropriate
when the unbroken group $H$ is the U(1) of electromagnetism.

The archetypical example \cite{hooft} of these occurs in an SU(2) theory
with a triplet Higgs field $\phi$ whose expectation value
$\langle \phi\rangle =v$ breaks the symmetry to U(1).  If the scalar
potential is
\begin{equation}
   V(\phi) = {\lambda \over 4} ( \phi^2 - v^2)^2
\end{equation}
the elementary excitations of the theory are a massless ``photon'', two
vector mesons $W^\pm$ with mass $m_W=ev$ and electric charges $\pm e$ (where
$e$ is the gauge coupling), and a neutral scalar with mass $m_H =
\sqrt{2\lambda} \, v$. 

Because $\Pi_2[{\rm SU(2)/U(1)}] = Z$, there is a single additive topological
charge $n$ that can take on any integer value.  It is related to the
magnetic charge by\footnote{In these units, the Dirac quantization
condition would require that $n$ be either integer or half-integer.
The more restrictive condition here follows from requiring that the
Higgs field be smooth at spatial infinity.  It is ultimately related
to the fact that one could add isospinor fields to the theory.  After
symmetry breaking, these would have electric charges $\pm e/2$.   The
existence of such charges would modify the Dirac condition so as to
require integer $n$.}
\begin{equation} 
     Q_M = {4 \pi n \over e}  \,\, .
\end{equation}
The classical solution corresponding to the unit monopole, with $Q_M=4\pi /e$,
is most often written in the spherically symmetric form 
\begin{eqnarray}
    A_i^a &=& \epsilon_{iak} {\hat r}_k  \left( { 1 -u(r) \over er}
    \right)  \cr 
    \phi^a &=& {\hat r}_a h(r) \,\, .
\label{radialgauge}
\end{eqnarray} 
(Here superscripts are SU(2) indices.)  Requiring that the solution be
nonsingular at the origin imposes the boundary conditions $u(0) =1$
and $h(0)=0$, while finiteness of the energy requires that $u(\infty)
=0$ and $h(\infty)=v$.  Substituting this ansatz into the field
equations leads to a pair of second order equations that can be solved
numerically.  The solution has a core region, of radius $R_{\rm
core} \sim 1/ev$.  Outside of this core $u$ and $v-h$ fall exponentially
fast and the field strength is, up to exponentially small terms, given
by 
\begin{equation}
      F_{ij}^a = \epsilon_{ijk} \hat r_a \hat r_i {1 \over er^2} \,\, .
\end{equation}
The factor of $\hat r_a$ indicates that in internal SU(2) space this
is parallel to $\phi$, and thus purely electromagnetic.  From the
remaining factors we see that this is just the Coulomb magnetic field
of a magnetic monopole with $Q_M=4\pi/e$.  The classical energy of the
solution, which gives the leading approximation to the monopole mass,
can be written as
\begin{equation}
    M = { 4 \pi v \over e} f(\lambda /e^2)
\end{equation}
where $f(s)$ increases monotonically between from 0 to 1.787
\cite{zachos} as $s$ varies from 0 to $\infty$.

Although the nontrivial topology of the Higgs field is most apparent
when the solution is written in the radial gauge form of
Eq.~(\ref{radialgauge}), a number of other aspects of the solution are
clarified by applying a singular gauge transformation that makes the
SU(2) orientation of the Higgs field uniform.  Choosing $\phi^a =
\delta^{a3} h(r)$, we may identify the third isospin component with
electromagnetism and write
\begin{eqnarray}
     A_i^{\rm EM} & \equiv& A_i^3 = {1 \over er} \left({ 1 - \cos\theta \over
\sin\theta} \right)   \cr
     W_i & \equiv& { A_i^1 + i A_i^2 \over \sqrt{2}} 
        = \hat n_i(\theta,\phi) {u(r) \over er} e^{i\alpha} \,\, .
\label{stringyform}
\end{eqnarray}
The first of these is just the U(1) vector potential for a point
monopole of magnetic charge $4\pi/e$, with the Dirac string
singularity lying along the negative $z$-axis.  In the second
equation, $\hat n_i$ is a complex unit vector whose precise form is
not needed for the present discussion.  This gives $u(r)/er$ a simple
interpretation as the magnitude of the massive $W$ field.  Thus, the
`t Hooft-Polyakov monopole is essentially a point Dirac monopole
surrounded by a core of massive vector field, with the components of
the latter arranged to give a singular magnetic dipole density that
cancels the singularity in the Coulomb magnetic energy
density.  Finally, note the arbitrary overall phase $\alpha$ that
appears in the expression for $W_i$.  This reflects the unbroken global
U(1) symmetry that survives even after the gauge has been fixed.

This monopole configuration is not an isolated solution, but rather
one of a many-parameter family of configurations, all with the same
energy.  Infinitesimal variations of these parameters correspond to
zero-eigenvalue modes in the spectrum of small fluctuations about the
monopole solution.  Three of these modes correspond to spatial
translations of the monopole; the corresponding parameters are most
naturally chosen to be the spatial coordinates of the center of the
monopole.  Because this is a gauge theory, there are also an infinite
number of zero modes that simply reflect the freedom to make local
gauge transformations.  To eliminate these, we must impose a gauge
condition.  However, this still leaves one zero mode, corresponding to
gauge transformations that do not vanish at infinity; roughly
speaking, this should be understood as a global gauge transformation in
the unbroken subgroup.  For example, if we impose the gauge conditions
that $\phi^1=\phi^2=0$ and that ${\bf A}^{\rm EM} = {\bf A}^3$ satisfy
the Coulomb gauge condition ${\bf \nabla} \cdot {\bf A}^{\rm EM}=0$,
the surviving gauge zero mode corresponds simply to shift of the U(1)
phase $\alpha$ in Eq.~(\ref{stringyform}).

When the system is quantized, a collective coordinate and a conjugate
momentum must be introduced for each of the zero modes.  For the
translation zero modes, the conjugate momentum is just the linear
momentum $\bf P$ of the monopole.  For the global gauge zero mode, the
collective coordinate is a U(1) phase variable, and the conjugate
momentum is the U(1) electric charge $Q_E$, which is proportional to the
time derivative of this phase; the quantization of $Q_E$ follows from
the periodicity of the U(1) phase.  Thus, allowing the collective
coordinates to vary linearly in time gives solutions with nonzero
momentum and electric charge and an energy of the form
\begin{equation}
   E = M + {{\bf P}^2 \over 2M} + {Q_E^2 \over 2 I} + O({\bf P}^4, Q_E^4)
\label{EwithPandQ}
\end{equation}
where the quantity $I$ can be expressed as a spatial integral
involving the square of the $W$ field.  The fact that the coefficient
of ${\bf P}^2$ turns out to be $1/2M$ is a nontrivial consequence of
the underlying Lorentz invariance of the theory.

\section{The BPS limit and electric-magnetic duality}

In general, the classical field equations governing the monopole can
only be solved numerically.  Matters are different, however, if one
works in the Bogomolny-Prasad-Sommerfield, or BPS, limit \cite{bps}.
Although initially introduced as a means of simplifying the classical
equations, it turns out to have a much deeper significance.

At the simplest level, the BPS limit is obtained by taking $\lambda$ and $m_H$
to zero with $v$ held fixed.   Since the field equations now depend only on a
single dimensionless constant that can be absorbed by a rescaling of fields
and distances, one might hope that they would become analytically tractable. 
Indeed, simply by guesswork one can find the solution   
\begin{eqnarray}
       u(r) &=& {v \over \sinh(evr) }         \cr
       h(r) &=& v \coth(evr) - {1\over er}     \,\, .
\label{bpssolution}
\end{eqnarray}
The $1/er$ tail of $h(r)$ reflects the fact that the scalar field
become massless in 
the BPS limit.  Such a massless field can mediate a long-range attractive
force, a fact that is of great importance.

Deeper insight is gained by examining the energy functional.  To simplify the
algebra, let us restrict ourselves to configurations with
vanishing electric charge and $A_0$ identically zero.  The energy for a static
configuration can then be written as\footnote{Here, and henceforth, I
write the gauge and Higgs fields as elements of the Lie algebra of the
group.  For the case of SU(2), these are related to the component
fields used in the previous section by $A_i = A_i^a \tau^a/2i$, where
the $\tau^a$ are the Pauli matrices.}
 \begin{eqnarray}
     E &=& \int d^3 x  \left[\frac{1}{2}\Tr B_i^2 +
    \frac{1}{2}  \Tr(D_i\phi)^2  + V(\phi) \right]     \cr \cr
       &=&  \int d^3 x  \left[\frac{1}{2} \Tr(B_i \mp D_i\phi)^2   
        \pm\Tr (D_i\phi) B_i + V(\phi)\right]
\label{BPSenergycalc}
\end{eqnarray}
where $B_i \equiv {1\over 2} \epsilon_{ijk} F_{jk}$.  The potential
term vanishes in the BPS limit.  A partial integration, together with
the Bianchi identity, gives
\begin{equation}
    \int d^3 x \Tr (D_i\phi) B_i =
     \int d^3 x \left[ \partial_i \Tr(\phi B_i) - \Tr\phi D_i B \right]
      =  \int d^2 S_i \Tr\phi B_i =  Q_M v   
\end{equation}
where the surface integral after the second equality is over a sphere
at spatial infinity.  Substituting this result back into
Eq.~(\ref{BPSenergycalc}) and using the upper (lower) sign for
positive (negative) magnetic charge gives the bound
\begin{equation}
      E \ge v |Q_M| = |n| \left({4 \pi v \over e}\right) \,\, .
\end{equation}
This bound is achieved by configurations that satisfy the first-order
equations
\begin{equation}
     B_i  = \pm D_i \phi \,\, .
\label{BPSeq}
\end{equation}
Because any static configuration that minimizes the energy is a
stationary point of the theory, solutions of Eq.~(\ref{BPSeq}) are
guaranteed to also be solutions of the full set of second-order field
equations; such solutions are referred to as BPS solutions.  For the
remainder of these lectures I will assume that the magnetic charge is
positive, and so will always use Eq.~(\ref{BPSeq}) with the positive
sign.

This argument is easily modified to include the possibility of dyons,
carrying nonzero electric charge $Q_E$ in addition to their magnetic charge. 
The BPS equations become 
\begin{eqnarray}
    B_i  &=& \cos \gamma \,D_i \phi     \cr
    E_i \equiv F_{0i} &=& \sin \gamma \,D_i \phi     \cr    
    D_0 \phi &=& 0 
\end{eqnarray}
where $\gamma = \tan^{-1}(Q_E/Q_M)$.   The solutions of these have energy
\begin{equation}
     E = v \sqrt{Q_M^2 + Q_E^2} \,\, .
\label{BPSdyonE}
\end{equation}
Although classically all values of $Q_E$ are possible, 
semiclassical quantization about the monopole shows that 
the electric charge must be an integral multiple of $e$.  Hence, for
weak coupling the lowest dyonic states have $Q_E \ll Q_M \sim 1/e$,
and Eq.~(\ref{BPSdyonE}) can be expanded to give
\begin{equation}
     E = v Q_M\left(1 + {Q_E^2 \over2 Q_M^2} + \cdots \right) 
\label{BPSdyonEexpansion}
\end{equation}
which is in agreement with Eq.~(\ref{EwithPandQ}).

The fact that the energy is strictly proportional to the magnetic
charge suggests that static multimonopole solutions might exist, since
the energy of such a solution would be just the sum of the masses of
the component monopoles and would be independent of their relative
positions.  From another point of view, this possibility, which is
actually realized, occurs because the magnetic repulsion between a
pair of static monopoles is exactly balanced by their mutual
attraction via the long-range scalar force.
 
A similar reduction of second-order field equations to a set of
first-order equations, with the energy of the solutions being
determined by the conserved charges, is found in a number of other
systems.  These include the Abelian Higgs model with parameters chosen
so that the scalar and vector masses are equal (corresponding to the
Type I-Type II boundary in superconductivity) \cite{abelianhiggs},
Chern-Simons-Higgs theory with a specially chosen scalar potential
\cite{CSvortices}, and the Yang-Mills instantons;
all of these admit static multisoliton solutions.

In all of these cases the theory can be trivially generalized to be
not only supersymmetric, but to also have extended supersymmetry.
Extended supersymmetry algebras include central charges.  These turn
out to be related to the charges carried by the solitons, so that the
BPS mass relations can be rewritten in terms of the generators of the
superalgebra.  For example, Eq.~(\ref{BPSdyonE}) is equivalent to
\begin{equation}
    H = \sqrt{ Z_1^2 + Z_2^2}
\label{ZtoH}
\end{equation}
where $H$ is the Hamiltonian and $Z_1$ and $Z_2$ are central charges.
The theory of supersymmetry representations shows that states
obeying such relations lie in ``short'' supermultiplets with fewer
components than otherwise.  (This generalizes the distinction between
massless and massive supermultiplets in $N=1$ supersymmetry.)  Since
small quantum corrections cannot change the size of a multiplet, these
mass relations must be exact and hold even when one passes from the
classical limit to the full quantum theory \cite{wittenolive}.
Furthermore, states obeying relations such as Eq.~(\ref{ZtoH}) are
invariant under a subset of the supersymmetry algebra.  At the
classical level, the vanishing of the corresponding supercharges
reduces to first-order equations such as Eq.~(\ref{BPSeq}).

In particular, the BPS monopole solutions arise naturally in the
context of $N=2$ or $N=4$ supersymmetric Yang-Mills theory.  In
addition to the gauge field, these contain $N/2$ Dirac fermion fields
and $2N-2$ real scalar fields $\phi_p$, all in the adjoint
representation.  The Lagrangian is
\begin{equation} 
   {\cal L} = \frac{1}{2}{\rm Tr} F_{\mu\nu}^2
     + \frac{1}{2}{\rm Tr} (D_\mu \phi_p)^2 
     -\frac{e^2}{4} {\rm Tr} \left( [\phi_p,\phi_q] \right)^2
     + {\rm fermion~terms} \,\, .
\end{equation}
If the $\phi_p$ with $p > 1$ are set identically to zero, the bosonic
parts of the classical field equations reduce to the BPS equations for
$A_\mu$ and $\phi_1$.

\begin{center}
\begin{tabular}{lcccccc}
&& \underline{Mass\vphantom{$Q_{E}$}} && \underline{$Q_E$}
&&\underline{$Q_M$} \\\\ 
photon && 0 && 0 && 0 \\\\
$\phi$ && 0 && 0 && 0 \\\\
$W^\pm$ && $ev$ && $\pm e$ && 0 \\\\
Monopole &&$ \high{4\pi v \over e}$ && 0 && $\pm \high {4\pi v \over e}$ 
\end{tabular}

\begin{quote}
{\bf Table 1:} {\small
The particle masses and charges in the BPS limit of the SU(2) theory.}
\end{quote}
\end{center}

Table~1 summarizes the masses and charges of particles of the
nonsupersymmetric SU(2) theory in the BPS limit. Montonen and Olive
\cite{dual} noted that this spectrum is invariant under the transformation
$$
   e \longleftrightarrow {4 \pi \over e} \qquad\qquad Q_E
   \longleftrightarrow Q_M  \,\, .\nonumber
$$
This led them to conjecture that the theory might be self-dual, with
the role of the solitons and of the elementary excitations being
interchanged under this transformation.  

An obvious difficulty with this conjecture is the fact that the
massive $W$'s have spin one, whereas the classical monopole and
antimonopole solutions are spherically symmetric and therefore lead to
spinless particles after quantization.  A resolution to this puzzle is
found when fermion fields are added to make the theory supersymmetric.
Each adjoint representation Dirac fermion field has two normalizable
zero modes in the presence of the singly charged monopole
\cite{jackiwrebbi}.  These lead to a degenerate multiplet of monopole
states that differ only in the occupation numbers of these modes.  The
theory with $N$-extended supersymmetry has $N/2$ Dirac fermions, hence
$N$ zero modes and $2^N$ degenerate states.  For $N=2$ these form a
spin-1/2 doublet and two spin-0 states, but not the spin-1 states that
are needed to match the $W$ bosons.  For $N=4$, there are 16 states,
including a spin-1 triplet, four spin-1/2 doublets, and five spin-0
states.  This not only gives the desired spin-1 states, but also
exactly matches \cite{osborn} the complete supermultiplet structure of
the electrically-charged elementary
excitations.\footnote{Supersymmetry is also essential for ensuring
that the spectrum of dyonic states is consistent with duality.  For
example, semiclassical quantization of the U(1) zero mode about the
unit monopole leads to a tower of states with unit magnetic charge but
multiple electric charge.  Duality then requires a tower of states
with unit electric charge and multiple magnetic charge.  Sen
\cite{sen} has shown how these can arise as zero energy bound states
in the supersymmetric theory; in Sec.~\ref{MSAsec} I will describe how
similar methods resolve some duality puzzles in a theory with a large
gauge group.}

Further support for the duality conjecture is obtained by considering
the low-energy scattering of electrically charged particles \cite{dual}.
The existence of classical static multimonopoles solutions in the BPS
limit was noted above.  The obvious dual to these would be states
containing several electrically charged elementary particles at rest.
Since we cannot construct a semiclassical approximation to these
states in the weak coupling limit, it is hard to study them directly.
Instead, we can examine the behavior of scattering states containing
two like-charged particles as their relative momentum goes to zero, and
look for the cancelation between the Higgs and electromagnetic forces
in the static limit.  As an example, consider the scattering of the
massive positively charged fermions that are the superpartners of the
massive gauge bosons.  There are two graphs that contribute at tree
level, one with a massless scalar exchanged between the fermions and
one where they exchange a photon.  (Note that the tree level
scattering amplitude is the same in the $N=2$ and $N=4$ theories.)
Let us work in the center of mass frame, with initial momenta
\begin{equation}
     p =  (E, {\bf P})       \qquad    k = (E, -{\bf P})  
\end{equation} 
and final momenta 
\begin{equation}
     p' =  (E, {\bf P}')       \qquad    k' = (E, -{\bf P}')   \,\, .
\end{equation} 
The contribution to the scattering amplitude
from scalar exchange is
\begin{equation} 
   {\cal M}_S =-{ie^2 \over q^2} \left[ \bar u(p') u(p) \right]
       \left[ \bar u(k') u(k) \right]
\label{MIcalc}
\end{equation}
where $q\equiv p'-p = k-k' = (0, {\bf P}' - {\bf P})$,
while that from photon exchange is 
\begin{equation}
    {\cal M}_V = {ie^2 \over q^2} g_{\mu\nu} 
       \left[ \bar u(p')\gamma^\mu u(p) \right]
       \left[ \bar u(k')\gamma^\nu  u(k) \right] \,\, .
\label{MIIcalc}
\end{equation}
Now use the fact that 
\begin{equation} 
    \bar u(p') \gamma^\mu u(p) = \bar u(p') \left[ {(p+p')^\mu \over 2m}
        + {i \sigma^{\mu\nu}q_\nu \over 2m} \right] u(p) \,\, .
\end{equation}
For $\mu = 1$, 2, or 3, the quantity sandwiched between the spinors is
manifestly of order $P$.  For $\mu =0$, if we write $E = m + {\bf
P}^2/2m + \cdots$ and recall that $\bar u(p') \sigma^{0i} u(p)$
vanishes if $\bf P=0$, we find that
\begin{equation} 
   \bar u(p') \gamma^0 u(p)  = \bar u(p') [ m + O({\bf P}^2) ] u(p) \,\, .
\end{equation}
After substituting this into Eq.~(\ref{MIIcalc}) and then combining
the result with Eq.~(\ref{MIcalc}), we see that the scattering
amplitude does indeed vanish in the static limit ${\bf P} \rightarrow
0$.

\section{SU(2) Multimonopole solutions and index theorems}

Let us return now to the classical SU(2) monopole solutions, focusing
in particular on those with higher magnetic charge, $n\ge 2$.  It was
argued previously that one might expect to find solutions
corresponding to several unit monopoles at rest.  One could also
imagine that there might be additional solutions, corresponding to new
species of monopoles with larger magnetic charges, that do not have
multiparticle interpretations.  A useful tool for exploring this
possibility is the use of index theory methods to count the
normalizable zero modes, and thus determine the number of collective
coordinates needed to characterize a solution \cite{SUtwoindex}.

Consider an arbitrary charge $n$ solution with gauge and Higgs fields
$A_i$ and $\phi$.  We are interested in small perturbations $\delta
A_i$ and $\delta\phi$ that preserve the BPS equation.  If we write
$A_i$ and $\phi$ as antihermitian matrices in the adjoint
representation of $SU(2)$ and $\delta A_i$ and $\delta\phi$ as
three-component column vectors, the equations resulting from the
variation of Eq.~(\ref{BPSeq}) take the form
\begin{eqnarray} 
    0 &=& \delta(B_j -D_j\phi)  \cr
      &=& D_j\,\delta\phi - e\phi \,\delta A_j - \epsilon_{jkl} D_k\, \delta
      A_l
\label{BPSpert}
\end{eqnarray}
where $D_j$ is the covariant derivative with respect to the
unperturbed gauge field.  We are not interested in zero modes that
simply correspond to local gauge transformations, although we do want
to keep those due to global gauge transformations, since excitations
of these give rise to electric charges.  To eliminate the unwanted
modes, we require that the perturbation be orthogonal to all local
gauge transformations, and hence satisfy
\begin{equation} 
    \int d^3x \left[ (\delta A_j)^t D_j \Lambda  + i e(\delta \phi)^t
           \phi \Lambda  \right]
\end{equation}
for any gauge function $\Lambda(x)$ that vanishes at spatial infinity.
By an integration by parts (valid only because $\Lambda$ vanishes
at large distance), this is equivalent to the background gauge
condition
\begin{equation}
      0 = D_j\,\delta A_j + e \phi \,\delta \phi \,\, .
\label{backgroundgauge}
\end{equation}
Our task is to determine the number of linearly independent normalizable
solutions of Eqs.~(\ref{BPSpert}) and (\ref{backgroundgauge}).

It is convenient to begin by defining \cite{lowell}
\begin{equation}  
    \psi = I \delta \phi + i \sigma_j \delta A_j  \,\, .
\end{equation}
This allows Eqs.~(\ref{BPSpert}) and (\ref{backgroundgauge}) to be
combined into a single Dirac equation
\begin{equation}
     0 = (-i \sigma_j D_j + ie \phi) \psi \equiv {\cal D} \psi  \,\, .
\label{BPSdiraceq}
\end{equation}
It should be kept in mind that solutions of Eq.~(\ref{BPSdiraceq}) that
differ by multiplication by $i$ correspond to linearly independent
solutions for $\delta A$ and $\delta \phi$, so the number of bosonic
zero modes is equal to twice the number of normalizable fermionic
solutions this equation.  Note also that Eq.~(\ref{BPSdiraceq}) is
preserved by transformations of the form 
\begin{equation}  
      \psi \rightarrow \psi e^{i{\bf v}\cdot \bsigma} \,\, .
\label{zeromodetrans}
\end{equation}
Given one bosonic zero mode, one can construct three more by means of
such transformations.  This relation among zero modes will be of
significance later.

Now define 
\begin{equation}
     {\cal I} = \lim_{M^2 \rightarrow 0}
     {\rm Tr}\, {M^2 \over {\cal D}^\dagger {\cal D}+M^2}
      - {\rm Tr}\, {M^2 \over {\cal D}{\cal D}^\dagger+M^2 } \,\, .
\label{calIdef}
\end{equation}
This counts the difference between the numbers of zeroes of the operators
${\cal D}^\dagger {\cal D}$ and ${\cal D} {\cal D}^\dagger$ or, equivalently,
the difference between the numbers of zeros of ${\cal D}$ and ${\cal
D}^\dagger$.  Because
\begin{equation}
    {\cal D} {\cal D}^\dagger =  - {\bf D}^2 + \phi^2
\end{equation}
is positive definite and has no normalizable zero modes, $2\cal I$
would seem to be precisely the quantity we want, with the factor of 2
taking into account the difference between the counting of bosonic and
fermionic modes.  However, the presence of massless particles in the
theory means that there will be a continuum spectrum with eigenvalues
extending down to zero.  If the continuum density of states were to be
sufficiently singular near zero, it could give a nonzero contribution
to $\cal I$.  A careful examination of the large distance behavior of
the various operators shows that this is not the
case.\footnote{Although the presence of the continuum spectrum does
not affect the correspondence between $\cal I$ and the counting of
normalizable zero modes, it does manifest itself by making the trace
in Eq.~(\ref{calIdef}) $M$-dependent; had the spectra of ${\cal
D}^\dagger {\cal D}$ and ${\cal D} {\cal D}^\dagger$ been purely
discrete, this trace would have been independent of $M$ and no limit
would have been needed.}  One can evaluate $\cal I$ by showing that it
can be written as the integral over all space of the divergence of a
current $J_i$.  Gauss's theorem converts this to a surface integral at
spatial infinity that is determined by the long-range behavior of the
Higgs and magnetic fields.  Evaluating this integral and taking the
limit $M^2 \rightarrow 0$ yields
\begin{equation}
     2 {\cal I} = 4n \,\, .
\end{equation}

Thus, any SU(2) BPS solution carrying $n$-units of magnetic charge
belongs to a $4n$-dimensional space of solutions, or moduli space.
The corresponding collective coordinates are just what would be
expected for a configuration of $n$ independent unit monopoles, each
with three position variables and one U(1) phase.  Allowing these to
become time-dependent would yield independent nonzero linear momenta
and electric charges for each of the individual monopoles.  Together
with the simple additive nature of the BPS mass formula, this strongly
suggests that all $n \ge 2$ classical solutions should be interpreted
as multimonopole solutions and that the corresponding states in the
quantum theory are multiparticle states.

\section{Fundamental monopoles in larger gauge groups}
\label{fundmonsec}

Let us now see what happens in the case of a larger simple gauge group
$G$, of rank $r> 1$, with an adjoint Higgs field $\phi$.  Recall that
the generators of a Lie algebra can be taken to be $r$ commuting
quantities $H_i$ that generate the Cartan subalgebra, together with a
raising or lowering operator $E_\balpha$ for each of the $r$-component
root vectors $\balpha$.  By making an appropriate choice of basis, any
element of the Lie algebra can be brought into the Cartan subalgebra.
In particular, we can choose to write the asymptotic Higgs field in
some reference direction to be of the form
\begin{equation}
   \phi_0 =   {\bf  h} \cdot {\bf H} \,\, .
\end{equation}
The $r$-component vector $\bf h$ determines the nature of the symmetry
breaking.  If it has nonzero inner product with all of the $\balpha$,
then $G$ is broken maximally, to ${\rm U}(1)^r$.  If instead there are 
some roots orthogonal to $\bf h$, then these form the root diagram for
some subgroup $K$ of rank $k$ and the unbroken gauge group is $K
\times {\rm U}(1)^{r-k}$.

Because the long-range part of the magnetic field must commute with
the asymptotic Higgs field, it too can be brought into the Cartan
subalgebra.  This allows us to define a second vector $\bf g$,
characterizing the magnetic charge, by requiring that the asymptotic
magnetic field in the direction used to define $\phi_0$ be of the form
\begin{equation} 
    B_k =  \frac{{\hat r}_ k}{ 4\pi r^2 } {\bf g}\cdot {\bf H } 
         +O(r^{-3}) \,\, .
\end{equation}
The topological quantization condition on the magnetic charge can then be
written as \cite{topology}
\begin{equation}
     e^{i{\bf g}\cdot {\bf H }} = 1 \,\, .
\end{equation}

Let us now concentrate on the case of maximal symmetry breaking,
returning later to the case where the unbroken symmetry contains a
non-Abelian factor.  To start, recall that a basis for the root
diagram of $G$ is given by $r$ simple roots $\bbeta_a$ with the
property that any other root can be written as a linear combination of
the $\bbeta_a$ with coefficients that are either all positive or all
negative.  The relative angles and lengths of the simple roots
characterize the Lie algebra (and are encoded in the Dynkin diagram),
but the choice of the simple roots is not unique, with the
various allowed choices being related by elements of the Weyl group.
However, the vector $\bf h$ uniquely determines a preferred set of
simple roots that satisfy the condition
\begin{equation} 
    {\bf h} \cdot     \bbeta_a \ge 0
\end{equation}
for all $a$.  The solution to the quantization condition can be written in
terms of these as
\begin{equation} 
    {\bf g} = \frac{4\pi}{e} \sum_{a=1}^r  n_a
          {{\mbox{\boldmath $\beta$}}}_a^*
\end{equation}
where the dual of a root is defined by $\balpha^* =
\balpha/\balpha^2$.  The coefficients $n_a$ are integers and are the
$r$ topological charges corresponding to the homotopy group
$\Pi_2[{\rm SU}(2)/{\rm U}(1)^r] = Z^r$.  For a BPS solution these
must all be of the same sign; without loss of generality, we can take
these to be positive, corresponding to the upper sign in
Eq.~(\ref{BPSeq}).

The BPS mass formula becomes
\begin{equation}
     M = \left({4 \pi \over e}\right)\sum_{a=1}^r n_a {\bf h} \cdot \bbeta_a
      \equiv \sum_{a=1}^r n_a m_a \,\, .
\end{equation}
An index calculation \cite{erick} similar to that for the SU(2) theory
shows that the 
number of normalizable zero modes is
\begin{equation} 
    2 {\cal I} = 4 \sum_{a=1}^r n_a \,\, .
\end{equation}
(Again, a detailed analysis shows that the continuum spectrum has no
effect on this result.)  These results suggest that any higher-charged
arbitrary solution should be viewed as a multimonopole solution
containing appropriate numbers of $r$ different species of fundamental
monopoles, with the $a$th species of fundamental monopole having
topological charges $n_b=\delta_{ab}$, mass $m_a$, and four collective
coordinates, three of which are position variables while the fourth is
a phase in the $a$th U(1).  In fact, one can easily construct the
classical solutions corresponding to these fundamental monopoles.  To
do this, note that each simple root $\bbeta_a$ defines an SU(2)
subgroup of $G$ with generators
\begin{eqnarray} 
t^1 &=& \frac{1 }{ \sqrt{2{{\mbox{\boldmath $\beta$}}_a^2}}} 
        \left(E_{{\mbox{\boldmath $\beta$}}_a} + 
        E_{-{\mbox{\boldmath $\beta$}}_a} \right) \nonumber \\
t^2 &=& -\frac{i}{ \sqrt{2{{\mbox{\boldmath $\beta$}}_a^2}}} 
        \left(E_{{\mbox{\boldmath $\beta$}}_a} - 
        E_{-{\mbox{\boldmath $\beta$}}_a} \right) \nonumber \\
t^3 &=&  {{\mbox{\boldmath $\beta$}}_a^*}  \cdot  {\bf H}  \,\, .
\label{SUtwoembed}
\end{eqnarray}
The corresponding fundamental monopole solution is obtained by
embedding the SU(2) unit mono\-pole solution, appropriately rescaled,
in this subgroup, and adding constant terms to the Higgs field so
that the asymptotic $\phi$ has the required eigenvalues.  
If $A_i^s({\bf r};v)$ and $\phi^s({\bf r};v)$ ($s=1,2,3$) are the
fields for the SU(2) solution corresponding to a Higgs expectation
value $v$, then the embedded solution is
\begin{eqnarray} 
A_i({\bf r}) &=& \sum_{s=1}^3 A_i^s({\bf r};\,{\bf h}
\cdot {\mbox{\boldmath $\beta$}}_a ) t^s   \nonumber\\
\phi({\bf r})&=& \sum_{s=1}^3 \phi^s({\bf r};\,{\bf h}
\cdot {\mbox{\boldmath $\beta$}}_a ) t^s 
 + ( {\bf h} - {\bf h}\cdot {{\mbox{\boldmath $\beta$}}_a^*} \,\,
{\mbox{\boldmath $\beta$}}_a)\cdot {\bf H}   \,\, .
\label{embed}
\end{eqnarray}
It is easily verified that this has mass $m_a$ and that there are
precisely four normalizable zero modes.

To make this more concrete, let us consider the case of SU(3) broken
to ${\rm U(1)}\times {\rm U(1)}$.  If the asymptotic Higgs field is
taken to be diagonal with its eigenvalues decreasing along the
diagonal, the SU(2) subgroups generated by the simple roots $\bbeta_1$
and $\bbeta_2$ lie in the upper left and lower right $2 \times 2$
blocks, respectively.  Embedding the SU(2) monopole in these subgroups
gives a pair of fundamental monopoles, each with four normalizable
zero modes.  The first has topological charges $n_1=1$ and $n_2=0$ and
mass $m_1$, while the second has $n_1=0$, $n_2=1$, and mass $m_2$. There is
also a third SU(2) subgroup, corresponding to the composite root
$\bbeta_1 + \bbeta_2$, lying in the four corner elements of the $3
\times 3$ SU(3) matrix.  Using this subgroup to embed the SU(2) unit
monopole gives a third solution which, like the previous two, is
spherically symmetric.  It has topological charges $n_1=n_2=1$ and
mass $m_1 +m_2$.  Unlike the two other embedding solutions, it has not
four, but instead eight, zero modes.  The extra modes correspond to
the fact that this solution can be continuously deformed into one
containing two widely separated fundamental monopoles, one of each
type.  Despite its spherical symmetry, it is just one of a family of
two-monopole solutions.  After quantization, these give rise to a set
of two-particle states, not to a new type of one-particle state.

Let us now consider these results in light of the Montonen-Olive
conjecture.  Table~2 lists the elementary vector particles, together
with the masses and the values of their electric-type charges under
the two unbroken U(1) factors.  (The remaining elementary
excitations are related to these by supersymmetry and follow the same
pattern.)  The two massless excitations, which carry no charge, should
be self-dual in the same sense as the photon of the ${\rm SU(2)} \rightarrow
{\rm U(1)}$ case.  The duals to the $W$-bosons corresponding to $\bbeta_1$
are clearly the states built upon the $\bbeta_1$-embedding of the
SU(2) monopole and antimonopole, and similarly for the $\bbeta_2$-$W$
boson.  What are the duals to the $W$'s corresponding to
$\bbeta_1+\bbeta_2$?  One might have thought that these would be
obtained from the $(\bbeta_1+\bbeta_2)$-embedding of the SU(2) solution.
However, we have just seen that this corresponds to a two-particle
state and so cannot give the dual to an electrically charged
single-particle state.  Instead the dual state must be some kind of
zero-energy bound state involving the two fundamental monopoles.  To
explore this possibility, we need to know more about the dynamics of
the monopoles.  A very useful tool for doing this is the moduli space
approximation.

\begin{center}
\begin{tabular}{lcccccc}
&&\underline{Mass\vphantom{$Q_{E1}$}} && \underline{$Q_{E1}$}
&&\underline{ $Q_{E2}$} \\\\ 
$\gamma_1$ && 0 && 0 && 0 \\\\
$\gamma_2$ && 0 && 0 && 0 \\\\
$\bbeta_1$-$W$ && $m_1$ && $\pm e$ && 0 \\\\
$\bbeta_2$-$W$ && $m_2$ && 0 && $\pm e$ \\\\
($\bbeta_1+\bbeta_2$)-$W$ && $m_1+m_2$ && $\pm $e && $\pm e$ \\\\
\end{tabular}

\begin{quote}
{\bf Table 2:} {\small
The particle masses and electric charges of the elementary vector
particles in 
the maximally broken SU(3) theory.  The two gauge bosons corresponding
to the unbroken generators are denoted by $\gamma_1$ and $\gamma_2$; 
$Q_{E1}$ and $Q_{E2}$ are the electric charges in the corresponding
U(1) subgroups.}
\end{quote}
\end{center}

\section{The moduli space approximation}
\label{MSAsec}
 
The essential assumption of the moduli space approximation
\cite{manton} is that if 
the monopoles are moving sufficiently slowly\footnote{In this context,
``slowly moving'' means not only that the spatial velocities must be
small, but also that the electric charges, which are proportional to
the time derivatives of the U(1) phases, must also be small.}, the
evolution of the field configurations can be approximated as motion on
the moduli space of BPS solutions. Thus, let $A^{\rm BPS}_a({\bf r},
z)$ be a complete family of gauge-inequivalent BPS solutions for a
given magnetic charge, with $a=1,2,3$ referring to the spatial
components of the gauge potential and $A_4 \equiv \phi$, while $z$
denotes the various collective coordinates.  In the $A_0=0$
gauge, the moduli space approximation amounts to assuming that the
field configuration at any time $t$ is of the form
\begin{eqnarray}
     A_a({\bf r}, t) = U^{-1}({\bf r}, t)\, A^{\rm BPS}_a({\bf r}, z(t))
         \,U({\bf r}, t)  
         -{i\over e} U^{-1}({\bf r}, t) \,\partial_a U({\bf r}, t)  \,\, .
\label{MSAfields}
\end{eqnarray}
(Here $\partial_4=0$.)  Differentiating with respect to time gives
\begin{equation}
    \dot A_a =  \dot z_j \left[{\partial A_a \over \partial z_j} + D_a
           \epsilon_j \right]  \equiv \dot z_j \, \delta_j A_a  
\end{equation}
where the infinitesimal gauge transformations generated by the
$\epsilon_j$ arise from differentiation of the factors of $U$.
Because they correspond to variations on the space of BPS solutions,
the $\delta_j A_a$ are zero modes about the monopole solution at a
given time.  The gauge functions $\epsilon_j$ are fixed by
Gauss's law, which much be imposed as a constraint when working in
$A_0=0$ gauge.  This gives
\begin{equation} 
    0 = -D_a F^{a0} = \dot z_j\, D_a \delta_j A_a 
\end{equation}
which implies that the $\delta_j A_a$ must obey the background gauge
condition, Eq.~(\ref{backgroundgauge}).

If we now substitute these results into the $A_0=0$ gauge field theory
Lagrangian, we obtain 
\begin{equation}
    L = {1\over 2} \int d^3r {\rm Tr}\left[ {\dot A_i}^2 + {\dot
          \phi}^2 + B_i^2 + 
          (D_i\phi)^2 \right] \,\, .
\end{equation}
With the fields given by Eq.~(\ref{MSAfields}), the integral of the last two
terms is just the energy of the static BPS solution.  This is
completely determined by the magnetic charge and is independent of the
$z_j$; i.e., it is a constant term having no effect on the dynamics.
The remaining terms are a quadratic form in
the $\dot z_j$, giving the moduli space Lagrangian
\begin{equation}
     L_{\rm MS} =  {1\over 2} g_{ij}(z) \dot z^i \dot z^j   + {\rm constant}
\label{MSAlagform}
\end{equation}
where  
\begin{equation}
    g_{ij}(z) = \int d^3r \left[ \delta_i A_k \,\,\delta_j A_k + 
        \delta_i \phi \,\, \delta_j \phi  \right] 
\label{MSAmetricdef}
\end{equation}
can be interpreted as a metric on the moduli space.  With this interpretation,
the solutions of the equations of motion are simply geodesic motions on the
moduli space.

We have thus reduced the field theory Lagrangian to one involving only
a finite number of degrees of freedom.  To make use of this reduced
Lagrangian, we need to determine the moduli
space metric.  There are at least three methods for doing this:

1.  If a complete family of BPS solutions $A^{\rm BPS}_a({\bf r}, z)$
is known explicitly, then one can solve for the background gauge
zero modes and then substitute these into Eq.~(\ref{MSAmetricdef}) to
obtain the metric.

2. In some cases, the mathematical conditions that the moduli space
metric must obey are so constraining as to essentially 
determine the metric.

3.  The original reason for introducing the moduli space approximation
was to obtain information about the low-energy interactions between
monopoles.  In some cases, this can be turned the other way around,
and knowledge of low-energy monopole dynamics can be used to infer the
moduli space metric.

The last of these strategies can be used to obtain the metric for the
portion of an $N$-monopole moduli space that corresponds to
widely-separated monopoles \cite{gary}.  Consider first two BPS
mono\-poles (or rather dyons) in the SU(2) theory, with positions ${\bf
x}_i$, spatial velocities ${\bf v}_i = \dot {\bf x}_i$, U(1) electric
charges $q_i$, and U(1) magnetic charges $g_i=4\pi/e$ ($i=1,2$).  In
order that the moduli space approximation be valid, the ${\bf v}_i$
and $q_i$ should be small.  When the separation between these is large
(i.e., $|{\bf x}_1 - {\bf x}_2| \gg m_W^{-1}$), the only nonnegligible
interactions between them are their mutual electromagnetic forces and
the long-range force scalar force mediated by the Higgs field.  The
effect of these on monopole 1 are described by the Lagrangian
\begin{eqnarray}
   L^{(1)}_{\rm SU(2)} &=& \sqrt{g_1^2 + q_1^2}\;\left|\phi_0 +
\Delta\phi^{(2)}({\bf x}_1) \right|
 \sqrt{1- {\bf v}_1^2}
   \,  +\, q_1 \left[ {\bf v}_1 \cdot {\bf A}^{(2)}({\bf x}_1)
    -  A_0^{(2)}({\bf x}_1)  \right]  \nonumber\\
    &+& g_1 \left[ {\bf v}_1 \cdot \tilde{\bf A}^{(2)}({\bf x}_1)
    -  \tilde A_0^{(2)}({\bf x}_1)  \right] \,\, .
\end{eqnarray}
The first term includes the effect of the Higgs field, which is
manifested through the effective reduction in the mass of monopole 1
due to the $1/r$ tail of the Higgs field of monopole 2, denoted here
by $\Delta \phi^{(2)}$.  The next term describes the effect on the
electric charge $q_1$ of moving in the vector potential ${\bf A}^{(2)}$
and scalar potential $A_0^{(2)}$ of monopole 2, while the last term
describes the effect on the magnetic charge $g_1$ of moving in the
dual potentials generated by monopole 2.  Expanding up to terms
quadratic in the ${\bf v}_i$ and $q_i$ gives
\begin{eqnarray}
   L^{(1)}_{\rm SU(2)} &=& -m_1 \left( 1 - \frac{1}{2}{\bf v}_1^2 +
\frac{q_1^2 }{2 g_1^2} \right) 
- \frac{g_1g_2 }{ 8\pi r_{12} } \left[({\bf v}_1 -{\bf v}_2)^2
- \left(\frac{q_1 }{g_1} - \frac{q_2}{ g_2}\right)^2 \right] \nonumber \\
&-&\frac{1}{ 4\pi } (g_1q_2 - g_2q_1) ({\bf v}_2 -{\bf v}_1)\cdot {\bf w}_{12}
\end{eqnarray}
where ${\bf w}_{12}$ denotes the Dirac vector potential at ${\bf x}_1$
due to a unit magnetic charge at ${\bf x}_2$.

Generalizing this to the case of many particles, but still in the SU(2)
theory, gives the many-particle Lagrangian
\begin{eqnarray}
    L_{\rm SU(2)}  = \frac{1}{2} M_{ij} \left( {\bf v}_i \cdot {\bf v}_j
      - {q_iq_j \over g^2}   \right)  + \frac{g }{4\pi} q_i
      {\bf W}_{ij}\cdot {\bf v}_j
\end{eqnarray}
where
\begin{equation}
 M_{ij} = \cases{ m  - \high \sum_{k\ne i}\high \frac{g^2}{ 4\pi r_{ik}} \, ,  
          \qquad i=j \cr
     \cr 
  \high \frac{g^2}{ 4\pi r_{ij}}\, , \qquad i\ne j}
\label{SUtwoM}
\end{equation} 
and
\begin{equation}
    {\bf W}_{ij} = \cases{ -\high \sum_{k\neq i}{\bf w}_{ik} \, ,
         \qquad i= j \cr 
         \cr
      {\bf w}_{ij} \, , \qquad i\ne j}
\label{SUtwoW}
\end{equation}
with ${\bf w}_{ij}$ being the value at ${\bf x}_i$ of the Dirac potential
due to the $j$th monopole.  
This is almost, but not quite, what we
need.  The moduli space Lagrangian is written in terms of the
generalized velocities $\dot z_j$.  However, the $q_j$ are momenta,
not velocities.  We must therefore introduce U(1) phase variables
$\xi_j$ conjugate to the $q_j$ and perform a Legendre transformation
\begin{eqnarray} 
     {\cal L}_{\rm SU(2)}({\bf x}_j, \xi_j) &=& L_{\rm SU(2)}({\bf x}_j, q_j)  
       + \sum_j \dot{\xi}_jq_j/e     \nonumber\\
    &=& \frac{1}{2} M_{ij}  {\bf v}_i \cdot {\bf v}_j +
    \frac{g^4}{2(4\pi)^2}(M^{-1})_{ij}
    \left( \dot\xi_i + {\bf W}_{ik}\cdot {\bf v}_k \right)
    \left( \dot\xi_j + {\bf W}_{jl}\cdot {\bf v}_l \right)    
\end{eqnarray}
to obtain a new Lagrangian that is of the form of
Eq.~(\ref{MSAlagform}).  From this we can immediately read off the
asymptotic metric describing widely-separated monopoles: 
\begin{equation}
    ds^2_{\rm asym} = 
       \frac{1}{2}M_{ij}d{\bf x}_i\cdot d{\bf x}_j+\frac{g^4}{2(4\pi)^2}
(M^{-1})_{ij}(d\xi_i+{\bf W}_{ik}\cdot d{\bf x}_k)(d\xi_j+{\bf W}_{jl}
\cdot d{\bf x}_l) \,\, .
\label{SUtwoMSAmetric}
\end{equation}

The extension of this analysis to the case of many widely-separated
fundamental monopoles in an arbitrary maximally broken gauge group is
surprisingly simple \cite{klee2}.  Each monopole is associated with a
fundamental root $\bbeta_i$, and has an electric charge $q_i$ and a
phase $\xi_i$ corresponding to the U(1) generated by $\bbeta_i \cdot
{\bf H}$.  The long-range forces between a pair of monopoles are all
proportional to the inner products of the corresponding simple roots.
Since the inner product of any two simple roots is always less than or
equal to zero, this means that the long-range interactions between two
different fundamental monopoles are either vanishing (if the monopoles
correspond to orthogonal simple roots), or else of the opposite sign
from those between two monopoles of the same species.  The only effect
on the metric is to insert factors of $\bbeta_i \cdot \bbeta_j$,
replacing Eqs.~(\ref{SUtwoM}) and (\ref{SUtwoW}) by
\begin{equation}
 M_{ij} = \cases{ m  - \high \sum_{k\ne i}\high  \frac{g^2 \bbeta_i^*\cdot
          \bbeta_k^*}{ 4\pi r_{ik}} \, ,   
          \qquad i=j \cr
     \cr 
  \high  \frac{g^2\bbeta_i^*\cdot \bbeta_j^*}{ 4\pi r_{ij}}\, , \qquad i\ne j}
\label{generalM}
\end{equation} 
and
\begin{equation}
    {\bf W}_{ij} = \cases{\high  -\sum_{k\neq i}\bbeta_i^*\cdot
         \bbeta_k^*{\bf w}_{ik} \, , 
         \qquad i= j \cr 
         \cr
      \bbeta_i^*\cdot \bbeta_j^*{\bf w}_{ij} \, , \qquad i\ne j \,\, .}
\label{generalW}
\end{equation}

Could this asymptotic form for the metric be in fact exact?  For the
case of two monopoles in the SU(2) theory, the answer is clearly no.
The matrix $M$ that appears in the asymptotic metric is
\begin{equation}
      M = \left(\matrix{m-\high {g^2 \over 4\pi r} & \high {g^2 \over
                4\pi r} \cr
                \high  {g^2 \over 4\pi r} & m -\high  {g^2 \over 4\pi
                r}}   \right) 
\end{equation}
where $m$ is the mass of a single monopole.  The determinant of $M$
vanishes, implying a singularity in the metric, at $r = g^2/(2\pi m) =
1/(2m_W)$.  It seems quite unlikely that the interactions of two
monopoles at such a distance would lead to such singular behavior.
Indeed, there are short-range interactions, due to the massive fields
in the monopole cores, that were not taken into account in the
asymptotic analysis.  If one works in the singular gauge of
Eq.~(\ref{stringyform}), these interactions can be characterized by the
gauge-invariant quantity\footnote{It is the deformations due to these
interactions that prevent solutions with two separated SU(2) monopoles
from being axial symmetric.}
${\rm Re}\,[{\bf W}_{(1)}^* \cdot {\bf W}_{(2)}]$.

Neither of these two objections applies in the case of a larger gauge
group, provided that the monopoles correspond to different fundamental
roots. 
There are no interactions at all, and the moduli space is flat, if the
roots are orthogonal, so we need only consider the case where $\bbeta_1
\cdot \bbeta_2 < 0$.  The presence of this negative inner product leads
to a sign change in $M$ and removes the singularity.  Furthermore, the
obvious generalization of the bilinear interaction term, ${\rm Tr}\,
{\bf W}_{(1)}^\dagger \cdot {\bf W}_{(2)}$, vanishes identically.  (The
vanishing of this bilinear is a consequence of the two unbroken U(1)
symmetries.)

While it thus seems possible that the asymptotic metric might be exact
in this case, further analysis is required to show that this is
actually so.  To begin, we need several properties of the moduli space
that hold whether or not the monopoles are of distinct species.  From
the counting of zero modes, we know that an $N$-monopole moduli space
is a $4N$-dimensional manifold.  Three of the coordinates can be taken
to be the position of the center-of-mass.  Because the center-of-mass
motion is constant and independent of the behavior of the other
variables, the moduli space can be factored as the product of a flat
three-dimensional space spanned by the center-of-mass coordinates and
a ($4N-5$)-dimensional manifold.  The latter can itself be factored,
at least locally, into the product of a flat $R^1$, corresponding to
an overall phase in the unbroken U(1) generated by ${\bf h}\cdot {\bf
H}$, and a ($4N-4$)-dimensional manifold depending only on relative
coordinates and phases.\footnote{Globally, the moduli space can be
written as the product of an $R^1$ times the ($4N-4$)-dimensional
relative manifold divided by a discrete normal subgroup.}
Furthermore, the relation between the zero modes expressed in
Eq.~(\ref{zeromodetrans}) implies a relationship between infinitesimal
motions on the moduli space that induces a quaternionic structure that
makes it a hyper-K\"ahler manifold.  (The hyper-K\"ahler property can
also be inferred from the extended supersymmetry that underlies the
BPS structure.)  Finally, the moduli space must have a rotational
isometry that reflects the rotational symmetry of physical space.

Taken together, these properties are rather restrictive.  This is
particularly the case for the two-monopole case, where the relative
moduli space is four-dimensional.  Any four-dimensional hyper-K\"ahler
manifold must be a self-dual Einstein space.  When combined with the
requirement of a rotational isometry, this leaves only four
possibilities:

1.  Four-dimensional Euclidean space

2.  The Eguchi-Hanson gravitational instanton \cite{eguchi}

3.  The Atiyah-Hitchin geometry \cite{atiyah}

4.  Taub-NUT space

The first of these can be ruled out, since it is flat and would imply
that there were no interactions between the monopoles.  The asymptotic
behavior of the second does not match that inferred from the
long-range monopole interactions, and so it too must be discarded.
The Atiyah-Hitchin manifold asymptotically approaches the behavior
described by Eq.~(\ref{SUtwoMSAmetric}), although it differs at short
distance and thus avoids the singularity at $r = g^2/(2\pi m)$; it is
the relative moduli space for two SU(2) monopoles \cite{atiyah} or,
more generally, for two identical monopoles with an gauge group.  The
only remaining possibility for the case of distinct fundamental
monopoles is Taub-NUT space.  Not only does this turn out to agree
asymptotically with the metric found above, but it is actually
identical to it for all values of the monopole separation
\cite{klee,gaunt,connell}.

The exact moduli space metric for than two SU(2) monopoles is not
known, although it is clear that it must deviate from the asymptotic
metric.  On the other hand, the asymptotic metric for many fundamental
monopoles, all corresponding to different simple roots of a large
group $G$, remains nonsingular for all values of the intermonopole
separations.  This fact led to the conjecture \cite{klee2} that the asymptotic
metric might also be exact for this case, a result that has since been
proven \cite{murray,chalmers}.

This discussion of the moduli space approximation was motivated by the
need to find a zero energy bound state that would provide the missing
state required by the Montonen-Olive duality conjecture.  Let us now
briefly return to this point.  As we have seen, the natural context
for this duality is $N=4$ extended supersymmetry.  This supersymmetry
must be taken into account when deriving a low-energy approximation
to the theory.  This leads to a generalized moduli space Lagrangian
that involves not only the bosonic coordinates $z^i$ that span the moduli
space, but fermionic coordinates $\psi^j$ as well.  For the $N=4$ case, one
obtains \cite{witten}
\begin{equation}
    L = {1\over 2} g_{ij}(z) \left[\dot z^i \dot z^j  
         +i \bar \psi^i \gamma^0  D_t \psi^j \right] 
          + {1 \over 6} R_{ijkl} (\bar \psi^i \psi^j)
	(\bar \psi^k \psi^l)
\end{equation}
where $R_{ijkl}$ is the Riemannian curvature on the moduli space and
$D_t$ is a covariant derivative.

When this theory is quantized, the bosonic coordinates are treated as
usual.  The fermionic coordinates give rise to a number of
discrete modes whose occupation numbers are either 0 or 1.  A state
can be described by a multicomponent wave function of the form
\begin{equation} 
     f^{(0)}(z) |\Omega\rangle + f^{(1)}_i(z) |\psi_i\rangle +
        f^{(2)}_{ij}(z) |\psi_{ij}\rangle + \cdots 
\end{equation}
where the labeling of the kets indicates the occupied fermionic
modes.  The antisymmetry of the fermionic variables implies that
$f^{(n)}(z)$ is antisymmetric in its indices and so can be viewed as
an $n$-form on the moduli space; a bound state corresponds to a
normalizable form.  The requirement that a state have zero energy
turns out to be equivalent to requiring that the forms in its wave
function be harmonic \cite{witten}.  The duality conjecture predicts a
single supermultiplet of zero energy bound states, and hence a single
normalizable harmonic form; since the dual of a harmonic form is also
harmonic, this form must be either self-dual or anti-self-dual.  Once
the explicit form of the two-monopole moduli space metric has been
determined, it turns out to be fairly straightforward to identify the
normalizable form required for self-duality in the maximally broken
SU(3) theory, and to show that this form is unique \cite{klee,gaunt}.

\section{Nonmaximal symmetry breaking}

Let us now return to the case of a nonmaximally broken symmetry, with
the unbroken subgroup being of the form $K\times {\rm U(1)}^{r-k}$,
where $K$ is a semisimple group of rank $k$.  As in the maximally
broken case, we choose a set of simple roots $\bbeta_a$ whose inner
products with $\bf h$ are nonnegative.  However, we now must
distinguish between the $r-k$ roots $\tilde \bbeta_i$ for which ${\bf
h}\cdot \tilde \bbeta_i >0$ and the remaining $k$ roots $\bgamma_i$
that are orthogonal to $\bf h$; the latter form a set of simple roots
for $K$.  Furthermore, this set of simple roots is not uniquely
determined.  Instead, there are several possible choices, all related
by Weyl reflections that correspond to gauge transformations of $K$.
Consider, for example, the case of SU(3) broken to ${\rm SU(2)}\times
{\rm U(1)}$, with the roots labeled as in Fig.~2 and the unbroken
SU(2) having roots $\pm \bbeta_2 = \pm \bgamma$.  The simple roots can
be chosen as in the maximally broken case discussed in
Sec.~\ref{fundmonsec}, with $\tilde \bbeta = \bbeta_1$ and $\bgamma
=\bbeta_2$, or they can instead be chosen to be $\tilde \bbeta' =
\bbeta_1 + \bbeta_2$ and $\bgamma'= - \bbeta_2$.  The two choices are
related by an $SU(2)$ gauge transformation.

\vskip 5mm
\begin{center}
\leavevmode
\epsfysize =2in\epsfbox{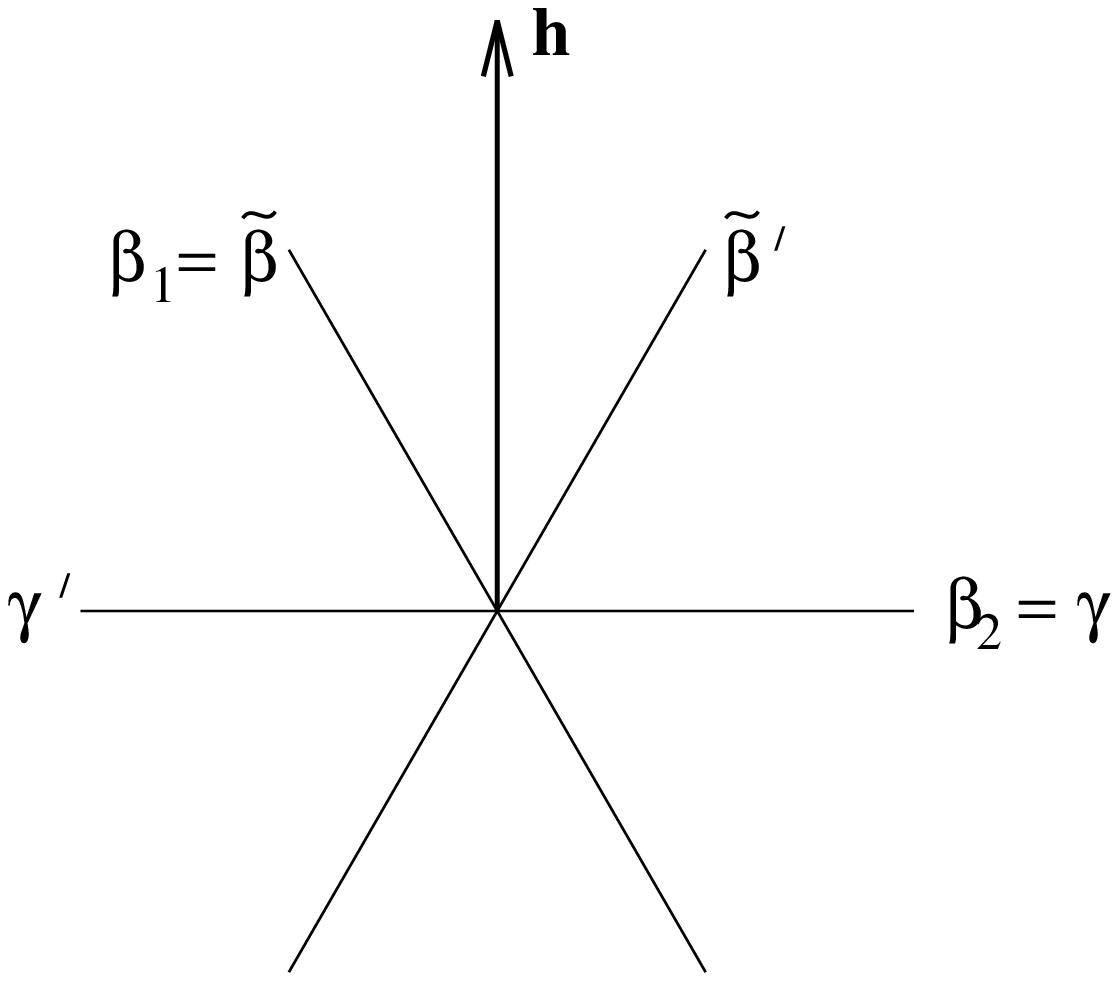}
\end{center}
\begin{quote}
{\bf Figure 2:} {\small
The root diagram of SU(3), with the Higgs vector $\bf h$ oriented to
give symmetry breaking to ${\rm SU}(2) \times {\rm U}(1)$.}
\end{quote}
\vskip 0.38cm

The quantization condition on the magnetic charge gives
\begin{equation} 
    {\bf g} = {4 \pi \over e} \left [ \sum_{a=1}^{r-k} \tilde n_a
          \tilde\bbeta_a^* + \sum_{j=1}^k q_k \bgamma_j \right] \,\, .
\label{nonmaximalg}
\end{equation}
The $r-k$ integers $\tilde n_a$ are the conserved topological
charges corresponding to the homotopy group $\Pi_2[G/(K\times {\rm
U(1)}^{r-k})]$.  They are gauge-invariant, and hence independent of
the choice of simple roots.  The $q_i$ are also integers, but are
neither gauge invariant nor conserved.  As before, the energy of a BPS
solution is determined by the 
magnetic charge, with
\begin{equation} 
     M =\left({4 \pi \over e}\right)\sum_{a=1}^{r-k}  \tilde n_a 
      {\bf h} \cdot \tilde \bbeta_a 
      \equiv \sum_{a=1}^{r-k} \tilde  n_a m_a \,\, .
\end{equation}
Note that only the topological charges $\tilde n_a$, and not the gauge-variant
$q_i$, appear in this formula.

In the maximally broken case we were able to identify $r$ species of
fundamental monopoles, each carrying one unit of a single topological
charge, that could be obtained by simple embeddings of the SU(2) unit
monopole.  Both the mass formula and the counting of zero modes
suggested that all BPS solutions should be interpreted as being
composed of an appropriate number of fundamental monopoles.  Can these
ideas be extended to the case with nonmaximal breaking?  Before
addressing this question, and indeed even before discussing the
counting of zero modes, it may be helpful to return to the SU(3)
example discussed in Sec.~\ref{fundmonsec}.

Recall that in the maximally broken case there were three spherically
symmetric solutions that could be obtained by SU(2) embeddings.  The
two fundamental monopoles, corresponding to the simple roots
$\bbeta_1$ and $\bbeta_2$, each had four zero modes.  The third
embedding solution, which corresponded to the composite root $\bbeta_1 +
\bbeta_2$, had a mass equal to the sum of the other two masses, had
eight zero modes, and was most naturally interpreted as a particularly
symmetric member of a family of two-monopole solutions.

Now let us consider these embedding solutions for the case of SU(3)
broken to SU(2)$\times$ U(1), with the unbroken SU(2) corresponding to
$\bbeta_2$ (i.e., lying in the lower left block).  As before, the
$\bbeta_1$-embedding gives a massive monopole solution.  Explicit
solution of the zero mode equations shows that there are precisely
four normalizable zero modes.  The $\bbeta_2$-embedding, on the other
hand, does not give a monopole solution.  Instead, Eq.~(\ref{embed}),
together with the fact that ${\bf h}\cdot \bbeta_2=0$, shows that this
embedding simple gives the vacuum.  Finally, the third embedding,
using $\bbeta_1 + \bbeta_2$, gives a solution that is gauge-equivalent
to the $\bbeta_1$-embedding, and hence also has four zero modes.

This last result is particularly puzzling if we think of the
nonmaximal breaking as a limiting case of maximal symmetry
breaking. How are eight zero modes suddenly converted into four?  One
way to understand this is to follow the behavior of the three
embedding solutions for the maximally broken case as ${\bf h}\cdot
\bbeta_2$ approaches zero.  The $\bbeta_1$-solution is independent of
${\bf h}\cdot \bbeta_2$, and neither it not its normalizable zero
modes are affected.  The $\bbeta_2$-solution exists (with four
normalizable zero modes) for any finite value of ${\bf h}\cdot
\bbeta_2$, but its core radius steadily increases, while the fields at
any fixed point tend toward their vacuum values, as ${\bf h}\cdot
\bbeta_2 \rightarrow 0$.  Although the mass and core radius of the
($\bbeta_1 + \bbeta_2$)-solution approach the values for the
$\bbeta_1$-monopole, nothing particularly dramatic happens to the
solution itself.  The zero modes are another matter.  Four of them
[the three translation modes and an overall U(1) mode] lie entirely
within the SU(2) subgroup defined by $\bbeta_1 + \bbeta_2$ and remain
normalizable.  The other four modes correspond to spatial separations
of the two component monopoles and hence grow in spatial extent as the
radius of the $\bbeta_2$-monopole increases; it is this growth, with
the consequent divergence at spatial infinity, that makes these modes
nonnormalizable in the limit of nonmaximal breaking.

How many zero modes would symmetry considerations have led us to
expect?  Usually, there is one zero mode for each symmetry of the
vacuum that is not a symmetry of the soliton.  Since the
$\bbeta_1$-monopole is not invariant under the unbroken SU(2), one
might at first expect to find three SU(2) modes in addition to the
three translational and one U(1) mode of the maximally broken
fundamental monopole, for a total of seven.  Because there is a U(1)
subgroup, generated by a linear combination of one SU(2) generator and
the original U(1) generator [i.e., by the $\lambda_8$ of the SU(3)],
that leaves the monopole invariant, this should be reduced to six.
However, it is clear that these SU(2) modes cannot be normalizable,
since an SU(2) rotation affects the $1/r$ tail of the vector
potential.\footnote{Transforming the modes into background gauge makes
the long-range behavior even worse \cite{abouel}.}

We can also use try to use index theory methods to count zero modes.
A generalization of the SU(2) calculation gives $2{\cal I}=6$ for
either the $\bbeta_1$- or the ($\bbeta_1 +\bbeta_2$)-monopole, in
agreement with the naive symmetry arguments.  However, $2{\cal I}$ is
equal to the number of normalizable zero modes only if the
contribution of the continuum spectrum to Eq.~(\ref{calIdef})
vanishes.  In the maximally broken case, analysis of the
large-distance behavior of the zero mode equations showed that the
continuum contribution vanished.  These arguments break down here,
essentially because of terms involving the $1/r$ tail of the vector
potential, thus allowing for a nonzero continuum contribution that
accounts for the discrepancy between $2 {\cal I}$ and the actual
number of normalizable modes.

The absence of normalizable zero modes associated with SU(2)
transformations means that there is no need to introduce collective
coordinates that specify the SU(2) orientation of the solution.  As a
result, the ``chromodyons'' --- solutions carrying SU(2) charges
associated with time-varying SU(2) collective coordinates --- that one
might have expected to find do not exist \cite{abouel}.  At a deeper
level, the absence of the chromodyons can be traced to the fact that
one cannot define ``global color''; i.e., there is a topological
obstruction to smoothly choosing a triplet of SU(2) generators over
the sphere at spatial infinity \cite{manohar}.

Thus, from many different approaches, we see anomalous aspects to
these solutions, always associated with the slow falloff of the
non-Abelian gauge fields at large distance.  This long-range tail is a
direct consequence of the fact that the SU(3) solutions that we have
considered all have magnetic charges with non-Abelian components;
i.e., their Coulomb magnetic fields are not invariant under the
unbroken SU(2).  This suggests that solutions whose magnetic charges
commute with the unbroken subgroup might be better behaved.

For an arbitrary gauge group broken to $K\times {\rm U(1)}^{r-k}$,
requiring that the magnetic charge commute with the unbroken subgroup
is equivalent to requiring that $\bf g$ be orthogonal to all the roots
of $K$,
\begin{equation}
     {\bf g}\cdot \bgamma_i =0 \,\, .
\label{purelyabelian}
\end{equation} 
Many of the anomalies described above are absent when this condition
holds.  The zero modes associated with the action of $K$ are
normalizable, and there is no obstruction to a global definition of
``$K$-color''.  The continuum contribution to $2{\cal I}$ vanishes, and
so index methods can be used to count the normalizable zero modes
\cite{erick2}.  One finds that there are
\begin{equation}
    2{\cal I} = 4  \left [ \sum_{a=1}^{r-k} \tilde n_a  
            + \sum_{j=1}^k q_k  \right]
\label{nonmaximalindex}
\end{equation} 
such modes.\footnote{As was noted above, the $q_i$ are gauge-variant,
and so the sum appearing in this equation is in general
gauge-invariant.  However, one can show that for magnetic charges
satisfying Eq.~(\ref{purelyabelian}) this sum, and hence the
expression for $2{\cal I}$, is gauge-invariant.  If the magnetic
charge does not obey Eq.~(\ref{purelyabelian}), this expression for
$2{\cal I}$ --- which in any case is no longer equal to the number of
normalizable zero modes --- is not valid.}

All this suggests that in studying the case of nonmaximal symmetry
breaking we should concentrate on configurations that obey
Eq.~(\ref{purelyabelian}) and thus have purely Abelian long-range
magnetic fields.  Imposing this constraint should not cause any
essential loss of generality, since any additional monopoles needed to
satisfy this condition can be placed arbitrarily far from the monopoles
of interest.  It also turns out to be helpful to treat this problem
as a limiting case of maximal symmetry breaking; i.e., to start with
maximal symmetry breaking and then consider the limit as some of the
${\bf h}\cdot \bbeta_a$ tend toward zero.  We will see that the
moduli space for the maximal broken case appears to behave smoothly in
this limit, and that the limiting value of the metric is indeed the
metric for the case of nonmaximal symmetry breaking.  Further, we will
see that even though some fundamental monopoles may become massless,
their degrees of freedom are not lost.

\section{An SO(5) example}

It is instructive to illustrate the case of nonmaximal symmetry
breaking with an example.  The simplest examples with unbroken
non-Abelian subgroups occur when a rank-two group $G$ is broken to
${\rm SU(2)} \times {\rm U(1)}$.  The choice $G={\rm SU(3)}$ is
perhaps the first that comes to mind.  With this choice,
Eq.~(\ref{purelyabelian}) requires that $\tilde n_1 = 2q_1$.  The
simplest possibility, $\tilde n_1=2$, $q_1=1$, would thus involve two
massive monopoles and have 12 zero modes.

A somewhat simpler example \cite{nonabelian} is obtained by taking the
gauge group to be SO(5), whose root diagram is shown in Fig.~3.  If
$\bf h$ is oriented as in Fig.~3a, the symmetry is maximally broken to
${\rm U(1)}\times {\rm U(1)}$, while if it is orthogonal to $\bgamma$,
as in Fig.~3b, the unbroken subgroup is ${\rm SU(2)} \times {\rm
U(1)}$.  Now consider the family of configurations with
\begin{equation}
    {\bf g} = {4 \pi \over e} \left(  \bbeta^* + \bgamma^* \right) \,\, .
\end{equation}
For nonmaximal symmetry breaking with $\bf h$ as in
Fig.~3b, these have ${\bf g} \cdot {\bf \bgamma}=0$, and
thus satisfy Eq.~(\ref{purelyabelian}).  According to
Eq.~(\ref{nonmaximalindex}), there is an 8-parameter family of
solutions.  As I will describe shortly, these solutions can be found
explicitly, and these solutions can be used to directly obtain the
moduli space metric.  If instead the symmetry breaking is maximal,
these solutions form a family of two-monopole solutions.  Since the
two component monopoles are distinct fundamental monopoles, the moduli
space metric is given by the results of Sec.~\ref{MSAsec}.  By taking
the limit ${\bf h}\cdot \bgamma \rightarrow 0$, we will be able to
compare the limit of the maximally broken metric with the metric
obtained directly from the solutions with nonmaximal breaking.

\begin{figure}
\begin{center}
\leavevmode
\epsfysize =2in\epsfbox{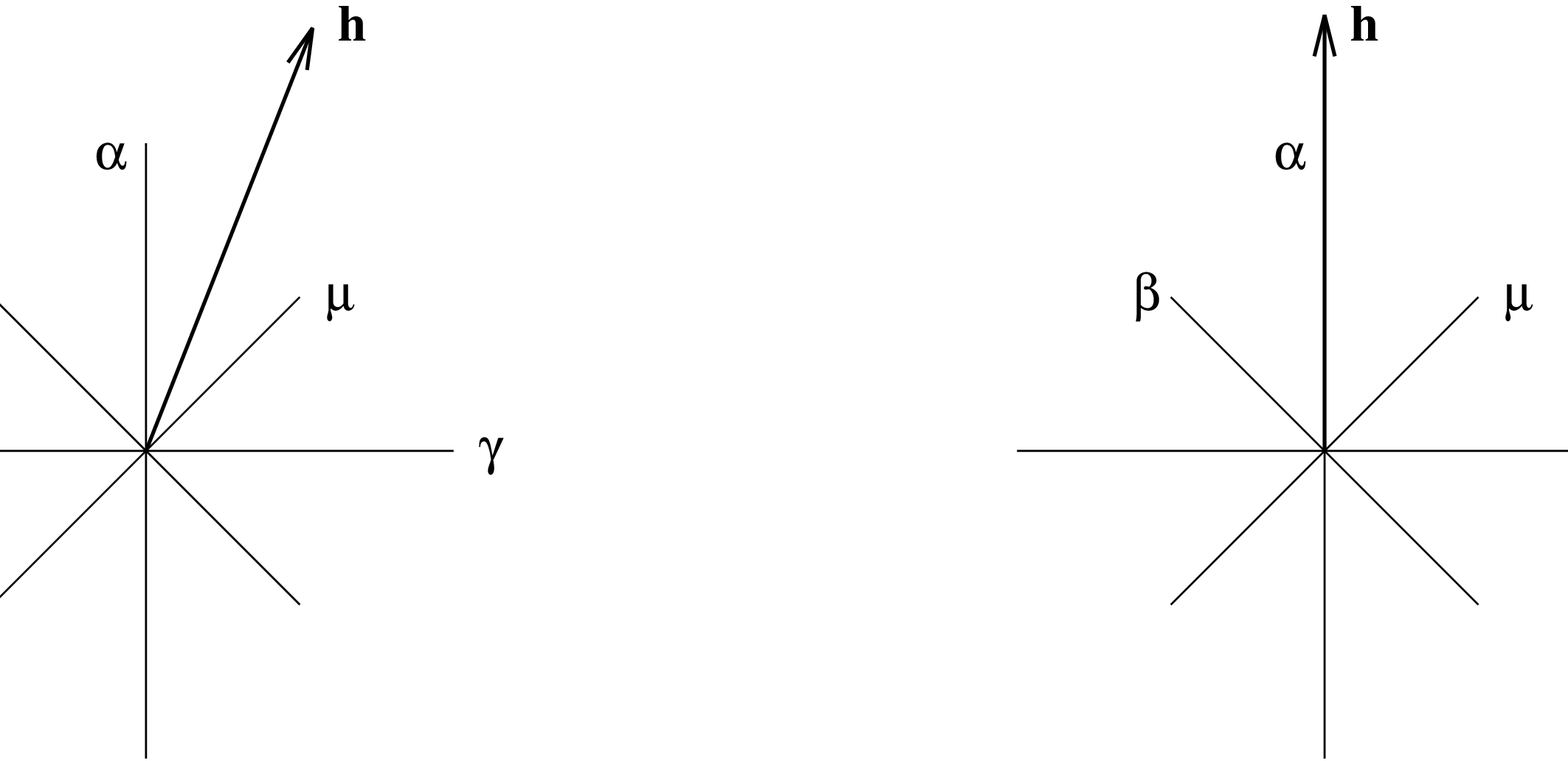}
\end{center}
\vskip 8mm
\begin{quote}
{\bf Figure 3:} {\small
The root diagram of SO(5). With the Higgs vector $\bf h$ oriented as
in (a) the gauge symmetry is broken to ${\rm U}(1)\times {\rm U}(1)$,
while with the orientation as in (b) the breaking is to ${\rm SU}(2)
\times {\rm U}(1)$.} 
\end{quote}
\end{figure}
\vskip 0.40cm

I will begin by describing the solutions for the case where SO(5) is
broken to ${\rm SU(2)} \times {\rm U(1)}$.  Of the eight parameters
entering these solutions, three clearly correspond to spatial
translation and just specify the position of the center of the
solution.  Four more must correspond to global ${\rm SU(2)} \times{\rm
U(1)}$ transformations.  This leaves only a single parameter whose
interpretation is not immediately obvious.  However, the solutions
must be spherically symmetric, since otherwise there would be at least
two rotational zero modes and hence two more parameters.  This
spherical symmetry can be used to reduce the field equations to a set
of ordinary differential equations that can be solved explicitly
\cite{so5}.  The extra parameter then appears as an integration constant
in this solution.

To write down the result, recall that the gauge field $A_i$ and the
Higgs field $\phi$ are both elements of the
10-dimensional Lie algebra of SO(5).  Any element $P$ of this Lie
algebra can be decomposed into a pair of three-component vectors ${\bf
P}_{(1)}$ and ${\bf P}_{(2)}$ and a $2 \times 2$ matrix $P_{(3)}$
obeying $P_{(3)}^* = - \tau_2 P_{(3)} \tau_2$ according to
\begin{equation}  
    P ={\bf P}_{(1)} \cdot {\bf t}({\mbox{\boldmath $\alpha$}}) 
       + {\bf P}_{(2)} \cdot {\bf t}({\mbox{\boldmath $\gamma$}}) 
       + {\rm tr}\, P_{(3)} {\cal M}
\end{equation}
where ${\bf t}(\balpha)$ and ${\bf t}(\bgamma)$ are defined by
Eq.~(\ref{SUtwoembed}) and
\begin{equation} 
       {\cal M} = \frac{i }{ \sqrt{{\mbox{\boldmath $\beta$}}^2}} 
       \left(\matrix{E_{\mbox{\boldmath $\beta$}}  & 
      -E_{-{\mbox{\boldmath $\mu$}}} \cr
       E_{\mbox{\boldmath $\mu$}}  & \,\,\,
       E_{-{\mbox{\boldmath $\beta$}}}} \right) \,\, .
\end{equation}
Under the unbroken SU(2) corresponding to $\bgamma$, ${\bf
P}_{(1)}$, ${\bf P}_{(2)}$, and  $P_{(3)}$ transform as three
singlets, a triplet, and a complex doublet, respectively.

Using this notation, the solution can be written as 
\begin{eqnarray}
   A^a_{i(1)} &=& \epsilon_{aim}{\hat r_m} A(r)  \qquad\qquad \quad 
   \phi^a_{(1)}= {\hat r_a} h(r) \nonumber \\
   A^a_{i(2)} &=& \epsilon_{aim}{\hat r_m} G(r,b) \qquad\qquad
   \phi^a_{(2)} = {\hat r_a} G(r,b) \nonumber \\
   A_{i(3)} &=& \tau_i F(r,b) \qquad\qquad \qquad \,\, 
   \phi_{(3)}= -i I F(r,b)   \,\, .
\label{myansatz}
\end{eqnarray}
The SU(2) singlet components $A^a_{i(1)}$ and $\phi^a_{(1)}$ are equal
to the monopole fields of Eq.~\ref{bpssolution}, with $A(r) =
[1-u(r)]/er$ and $v = {\bf h}\cdot \balpha$, while
\begin{eqnarray} 
   F(r) &=& { v \over \sqrt{8} \cosh (evr/2) }  L(r, b)^{1/2} \\
        \nonumber \\
   G(r) &=& A(r) L(r, b)
\end{eqnarray}
with 
\begin{equation}
   L(r,b)= \left[ 1 +  (r/b) \coth(evr/2) \right]^{-1}  \,\, .
\end{equation}

The quantity $b$ that enters through the function $L$ is the new
parameter; it can take on any positive real value.  Its significance
can be seen by examining the large-distance behavior of the solution.
As with the SU(2) monopole, there is a core of radius $\sim 1/ev$.
Outside this core, $A(r)$ falls as $1/r$, producing a $1/r^2$ Coulomb
magnetic field in the unbroken U(1), while $F(r)$ falls exponentially.
The behavior of $G(r)$ depends on the relative size of $r$ and $b$.
Because 
\begin{equation} 
      L \approx  \cases{ 1 \, ,\qquad 1/ev \simle r \simle b \cr
                       \high  {b\over r}  \, ,\qquad r \simge b}
\end{equation} 
$G(r)$ falls as $1/r$ when $1/ev \simle r \simle b$.  This corresponds
to a $1/r^2$ magnetic field in the unbroken SU(2), so in this region
the solution appears to carry both U(1) and SU(2) magnetic charge.
However, the $1/r$ falloff of $L$ for $r \simge b$ implies that the
SU(2) component of the magnetic field must fall at least as fast as
$1/r^3$ at large distance, so in actuality there is only a U(1)
magnetic charge.  Thus, one can think of these solutions as being
composed of a massive $\bbeta$-monopole, with a core of radius $\sim
1/ev$, surrounded by a ``cloud'' of non-Abelian fields of radius $b$.
The eighth, nonsymmetry-related, zero mode corresponds to the fact
that the energy of these solutions is independent of the value of the
``cloud parameter'' $b$.

The metric for the eight-dimensional moduli space of these solutions
is easily obtained.  First, the terms corresponding to the translation
and U(1) coordinates follow immediately from the mass formula for a
dyon.  Next, the $b$-zero mode can be obtained by variation of the
explicit solutions.  Because this zero mode turns out to already
satisfy the background gauge condition, it can be immediately
substituted into Eq.~(\ref{MSAmetricdef}) to give $g_{bb}$.  Finally,
by proceeding as described below Eq.~(\ref{zeromodetrans}), one can
obtain three more modes from this zero mode.  These correspond to
global SU(2) transformations and, after conversion to the standard
normalization, give the remaining components of the metric.  The
result is that
\begin{equation}
    ds_{SU(2)\times U(1)}^2 = M d{\bf x}^2 +  {16\pi^2 \over M }d\chi^2 + 
       k \left[{ db^2 \over b} 
      + b \left( d\alpha^2 + \sin^2\alpha\, d\beta^2  
        + (d \gamma + \cos\alpha\, d\beta)^2 \right) \right]  
\label{SOfivemetric}
\end{equation}
where $M$ is the total mass of the solution, $\alpha$, $\beta$, and
$\gamma$ are SU(2) Euler angles, $\chi$ is a U(1) phase angle, and $k$
is a normalization constant whose value is not important for our
purposes.

This should be compared with the metric for moduli space of solutions
with one massive $\bbeta$-monopole and one massive
$\bgamma$-monopole in the maximally broken case.
Equations~(\ref{SUtwoMSAmetric}--\ref{generalW}) lead to
\begin{eqnarray}
    ds_{U(1)\times U(1)}^2 &=& M d{\bf x}_{\rm cm}^2 +  {16\pi^2 \over M
    }d\chi_{\rm 
    tot}^2 +  \left(\mu +{k\over r}\right) \left[ dr^2 +r^2(d\theta^2
    + \sin^2\theta\, d\phi^2) \right]  \cr & & \qquad 
  + k^2 \left(\mu +{k\over r}\right)^{-1} (d\psi +\cos\theta \,d\phi)^2 \,\, .
\label{SOfiveMSBmetric}
\end{eqnarray}
Here $M$ is the sum of the two monopole masses, $\mu$ is the reduced
mass, and $r$, $\theta$, and $\phi$ specify the relative position vector
${\bf r} = {\bf r}_1 - {\bf r}_2$.  The overall U(1) phase
(corresponding to the subgroup generated by ${\bf h}\cdot {\bf H}$)
is $\chi$, while the relative U(1) phase is $\psi$.   Finally, $k$ is
the same constant as in Eq.~(\ref{SOfivemetric}).  

Nonmaximal breaking corresponds to the limit ${\bf h}\cdot \bgamma
\rightarrow 0$ in which the $\bgamma$-monopole becomes massless and
the reduced mass $\mu$ vanishes.  Setting $\mu=0$ in
Eq.~(\ref{SOfiveMSBmetric}) gives
\begin{equation}
     ds_{U(1)\times U(1)}^2 =  M d{\bf x}_{\rm cm}^2 +  {16\pi^2 \over M
          }d\chi_{\rm  tot}^2 
         +  k \left[{ dr^2 \over r} 
      + r \left( d\theta^2 + \sin^2\theta \,d\phi^2  
        + (d\psi + \cos\theta\, d\phi)^2 \right) \right] \,\, .
\end{equation}
This is exactly the same as Eq.~(\ref{SOfivemetric}), except for the
change in notation \goodbreak 
\begin{eqnarray}
           r  &\longleftrightarrow&  b   \nonumber\cr
           \theta  &\longleftrightarrow&  \alpha   \nonumber\cr
           \phi  &\longleftrightarrow&  \beta   \nonumber\cr
           \psi  &\longleftrightarrow&  \gamma
\end{eqnarray} 

Thus, at the level of the moduli space Lagrangian, the degrees of
freedom of the $\bgamma$-monopole survive even as the monopole becomes
massless in the ${\bf h}\cdot \bgamma \rightarrow 0$ limit.  However,
the relation of these degrees of freedom to the classical solution
changes.  First, as its mass $m_\bgamma$ decreases, the core of the
$\bgamma$-monopole spreads out until it becomes a cloud surrounding
the $\bbeta$-monopole, with the intermonopole separation becoming the
cloud size.  Second, the directional angles of the $\bbeta$-monopole
combine with the relative phase to give the Euler angles specifying
the SU(2) orientation of the cloud.  A consequence of the latter fact
is a certain ambiguity in the position of the massless monopole:
Initial configurations of massive monopoles that differ only in the
relative direction of the two monopoles become gauge-equivalent in the
$m_\bgamma\rightarrow 0$ limit,

\section{More complex examples with massless monopoles}

We can gain further insight into the meaning of these massless
monopoles by considering some more complicated examples.  In
particular, let us consider the case of ${\rm SU}(N)$ broken to ${\rm
U(1)} \times {\rm SU}(N-2) \times {\rm U(1)}$, with the unbroken ${\rm
SU}(N-2)$ corresponding to the middle $N-3$ roots of the Dynkin
diagram in Fig.~4.  As before, we are interested in solutions that
satisfy Eq.~(\ref{purelyabelian}), so that their asymptotic magnetic
fields are purely Abelian.  All solutions of this equation can be
written as sums of the following irreducible solutions:

1) $N-2$ massive and $(N-2)(N-3)/2$ massless monopoles, with 
\begin{eqnarray}
    \tilde n_1 &=& 0  \cr
    \tilde n_2 &=& N-2  \cr
      q_j &=& j \, , \qquad j=1,2, \dots, N-3
\end{eqnarray}

2) $N-2$ massive and $(N-2)(N-3)/2$ massless monopoles, with 
\begin{eqnarray}
    \tilde n_1 &=& N-2 \cr
    \tilde n_2 &=& 0  \cr
      q_j &=& N-2-j \, , \qquad j=1,2, \dots, N-3
\end{eqnarray} 

3)  Two massive and $N-3$  massless monopoles, with 
\begin{eqnarray}
    \tilde n_1 &=& 1 \cr
    \tilde n_2 &=& 1 \cr
      q_j &=& 1 \, , \qquad j=1,2, \dots, N-3
\end{eqnarray} 

\vskip 5mm
\begin{center}
\leavevmode
\epsfysize =0.6in \epsfbox{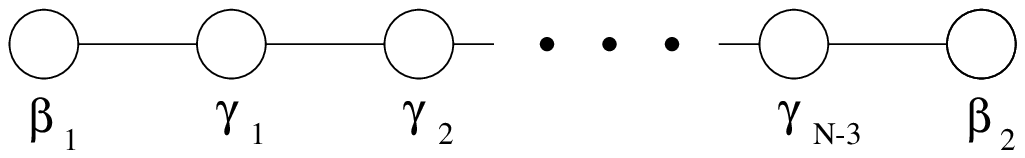}
\end{center}
\begin{quote}
{\bf Figure 4:} {\small 
The Dynkin diagram of SU(N), with the labeling of the simple roots
corresponding to symmetry breaking to ${\rm U}(1)\times {\rm SU}(N-2)
\times {\rm U}(1)$. }
\end{quote}
\vskip 0.38cm

In each of the first two cases Eq.~(\ref{nonmaximalindex}) gives a
total of $2(N-2)(N-1)$ zero modes.  The positions and U(1) phases of
the massive monopoles account for $4(N-2)$ of these, while global
${\rm SU}(N-2)$ transformations give $(N-2)^2 -1 $ more.  This leaves
$(N-3)^2$ modes that presumably correspond to parameters describing
gauge-invariant aspects of the non-Abelian cloud, showing that the
structure of these clouds can be much more complex than in the simple
SO(5) example of the previous section.  It would be enormously
instructive to examine these solutions in detail.  Unfortunately,
these solutions are not yet known explicitly, although there has been
considerable progress \cite{irwin} for the simplest case ($N=4$),
which can be viewed as ${\rm SU}(3) \rightarrow {\rm SU(2)} \times
{\rm U(1)}$.

The third case turns out to be more tractable \cite{nahmpaper}.  Here
there are $4(N-1)$ parameters, of which eight specify the positions
and U(1) phases of the two massive monopoles.  The number of remaining
parameters is clearly less than the dimension of ${\rm SU}(N-2)$.
This is explained by the fact that any solution with this magnetic
charge can be written as an embedding of an ${\rm SU}(4) \rightarrow
{\rm U(1)} \times {\rm SU(2)} \times {\rm U(1)}$ solution.  Hence,
there is a ${\rm U}(N-4)$ subgroup of the unbroken group that leaves
any given solution invariant, so the number of global gauge zero modes
is $4N-13$, which is the dimension of ${\rm SU}(N-2)/{\rm U}(N-4)$.
The one remaining zero mode corresponds to the single cloud parameter,
which I will again denote by $b$.

Even in this last case, the solutions are too complicated to be found
by a direct attack on the field equations.  However, they can be
obtained by making use of a construction due to Nahm \cite{nahm}.  In
this construction, one begins with a triplet of matrices $T_i(s)$
(where $s$ is a real variable associated with the eigenvalues of the
asymptotic Higgs fields) that obey a nonlinear differential equation.
These are then use to construct a linear equation that is obeyed by a
quantity $v({\bf r},s)$.  Finally, the gauge and Higgs fields for the
monopole solution are obtained from integrals involving $v$ and its
first derivative.

To be explicit, consider an SU($N$) theory where the asymptotic Higgs
field is of the form
\begin{equation}
   \phi= {\rm diag}\,(s_1, s_2, \dots , s_N)
\end{equation}
with $s_1 \le s_2 \le \cdots, \le s_N$, while the magnetic charge is
\begin{equation}
   Q_M= {4\pi \over e} {\rm diag}\,(n_1 , n_2-n_1, n_3-n_2, \dots,
   -n_{N-1})   \,\, .
\end{equation}
Define the stepwise continuous function $k(s)$ to be equal to $n_j$ on
the interval $n_j < s < n_{j+1}$.  The $T_i$ are required to have
dimension $k(s) \times k(s)$, and to satisfy\footnote{For the
remainder of this section, I will set $e=1$.}
\begin{equation}
   {dT_i \over ds} = {i \over 2} \epsilon_{ijk} [T_j, T_k]
       + \sum_P (\alpha_P)_i \delta(s-s_P) \,\, .
\label{Nahmeq}
\end{equation}
Here $s_P$ denotes any of the points $s_j$ such that $n_{j-1}=n_j$,
and the $(\alpha_P)_i$ are constant matrices of dimension 
$k(s_P) \times k(s_P)$.

The matrix $v({\bf r},s)$ is of dimension $2k(s) \times N$, and obeys
\begin{equation}
   0 =  \left[ -{d\over ds} + (T_I +r_i I) \otimes \sigma_i \right] 
        v({\bf r},s) + \sum_P a^\dagger_P S_P({\bf r})\delta(s-s_P)
\label{vequation}
\end{equation}
together with the normalization condition
\begin{equation}
     I =  \int ds\, v^\dagger({\bf r},s)v({\bf r},s) 
         + \sum_P S^\dagger_P({\bf r}) S_P({\bf r})  \,\, .
\label{vnormalization}
\end{equation}
Here $S_P({\bf r})$ is an $N$-component row vector, while
$a^\dagger_P$ is a $2k(s_P)$-component column vector with the property
that $a^\dagger_P a_P = (\alpha_P)_i \sigma_i - i (\alpha_P)_0 I$ has
rank one.\footnote{These two equations are preserved if $v$ and the
$S_P$ are multiplied on the right by $N\times N$ unitary matrices;
such transformations correspond to gauge transformations of the
spacetime fields. }

Finally, the gauge and Higgs fields for the monopole solution are
obtained from 
\begin{eqnarray}
    {\bf A}({\bf r}) &=& i \int ds\, v^\dagger({\bf r},s) {\bf \nabla}
               v({\bf r},s) 
        + \sum_P S^\dagger_P({\bf r}) {\bf \nabla}S_P({\bf r}) 
        \cr
    \phi({\bf r}) &=&  \int ds\, s \, v^\dagger({\bf r},s)v({\bf r},s)
      + \sum_P s_P S^\dagger_P({\bf r}) S_P({\bf r})  \,\, .
\label{Nahmfields}
\end{eqnarray}

In our case, the $n_j$ are all equal to unity and so the $T_i(s)$ are
all $1 \times 1$.  Equation~(\ref{Nahmeq}) reduces to a trivial equation
whose solution is that the $T_i(s)$ form a piecewise constant vector
\begin{equation}
      {\bf T}(s) = -{\bf x}_a \, , \quad s_a < s < s_{a+1} \,\, .
\end{equation}
With the $s_j$ all different, so that the symmetry
breaking is maximal, it is clear that the ${\bf x}_a$ should be
interpreted as the positions of the $N-1$ massive monopoles.  

With the SO(5) example in mind, one might expect to find some
ambiguity in the positions of the ``massless monopoles'' when the
middle $N-2$ eigenvalues of $\phi$ are set equal to give an unbroken
${\rm U(1)} \times {\rm SU}(N-2) \times {\rm U(1)}$.  Tracing through
the steps that lead from the $T_i$ to the space-time fields ${\bf A}$
and $\phi$, one finds that this is indeed the case.  The space-time
fields are unaffected by any transformation of the massless monopoles
positions ${\bf x}_2$, ${\bf x}_3$, \dots, ${\bf x}_{N-2}$ that
leaves invariant the 
sum of distances
\begin{equation}
    \sum_{a=1}^{N-2} \left|{\bf x}_{a+1} - {\bf x}_a \right| \equiv
         2b + \left|{\bf x}_{N-1} - {\bf x}_1 \right| = 2b +R \,\, .
\end{equation}
Just as in the SO(5) example, all but one of the massless monopole
coordinates are transformed into gauge-orientation parameters, leaving
only a single gauge-invariant cloud parameter $b$.

Because of the particularly simple form of the $T_i$ here,
Eq.~(\ref{vequation}) can be solved in closed form.  After rescaling
the solution so that the normalization condition
Eq.~(\ref{vnormalization}) is satisfied, it is straightforward,
although perhaps a bit tedious, to substitute the result into
Eq.~(\ref{Nahmfields}) and then integrate to obtain explicit
expressions for ${\bf A}$ and $\phi$; these involve only rational and
hyperbolic functions.  Because of the lack of spherical symmetry, these
expressions are naturally more complex than in the SO(5) case.
Nevertheless, they have some simplifying features that are reminiscent
of the SO(5) solution.  The fields can be decomposed into three pieces
that correspond (at least at large distances) to terms
transforming  under the singlet,
the fundamental, and the adjoint representations of the unbroken ${\rm
SU}(N-2)$.  Only the latter two depend on $b$, and in both cases this
dependence is through a single function $L$, which is now a matrix.
In addition, the Higgs fields for the fundamental and adjoint pieces
are given in terms of the same spacetime functions as the gauge
fields.

In fact, it is possible to choose the gauge so that, apart from a
constant contribution to $\phi$, all nonzero terms in these fields lie
in a $4 \times 4$ block.  This shows that the solution is essentially
an embedded SU(4) solution, and thus invariant under a U($N-4$)
subgroup, as was claimed above.  

\def\bhyl{\hat{\bf y}_L}
\def\bhyr{\hat{\bf y}_R}

The details of these solutions inside the massive monopole cores are
not very illuminating.  On the other hand, insight into the nature of
the non-Abelian cloud can be obtained by examining the asymptotic
behavior of the fields well outside the cores.  I will display the
form that these take for the SU(4) case; the extension to 
$N>4$ is straightforward.

Consider first the case $b \gg R$.  If the distances $y_L$ and $y_R$
from a point $\bf r$ to the two massive monopoles are both much less
than $b$, the Higgs field and magnetic field can be written in the
form
\begin{equation}
    \phi({\bf r}) = U_1^{-1}({\bf r})
        \left( \matrix{  t_4  - \high {1\over 2y_R} & 0 & 0 & 0 \cr\cr
             0 & t_2+ \high {1\over 2y_R} & 0 & 0 \cr\cr
             0 & 0 & t_2 - \high {1\over 2y_L} & 0 \cr\cr
             0 & 0 & 0 & t_1 + \high {1\over 2y_L} } \right) 
        U_1({\bf r})    + \cdots
\end{equation}
\begin{equation}
    {\bf B}({\bf r}) = U_1^{-1}({\bf r})
        \left( \matrix{ \high  {\bhyr\over 2y_R^2} & 0 & 0 & 0 \cr\cr
             0 &  - \high {\bhyr\over 2y_R^2} & 0 & 0 \cr\cr
             0 & 0 &\high{ \bhyl\over 2y_L^2} & 0 \cr\cr
             0 & 0 & 0 &  - \high {\bhyl\over 2y_L^2}   } \right) 
        U_1({\bf r}) + \cdots
\end{equation}
where $U_1({\bf r})$ is an element of SU(4) and the dots represent
terms that are suppressed by powers of $R/b$, $y_L/b$, or $y_R/b$.
These are the fields that one would expect for two massive monopoles,
each of whose magnetic charges has both a U(1) component and a
component in the unbroken SU(2) that corresponds to the middle $2
\times 2$ block.  If instead $y \equiv (y_L +y_R)/2 \gg b$,
\begin{equation}
    \phi({\bf r}) = U_2^{-1}({\bf r})
        \left( \matrix{ t_4 - \high {1\over 2y} & 0 & 0 & 0 \cr\cr
             0 & t_2 & 0 & 0 \cr\cr
             0 & 0 & t_2  & 0 \cr\cr
             0 & 0 & 0 & t_1 + \high {1\over 2y}  } \right) 
        U_2({\bf r})    + O(b/y^2)
\end{equation}
\begin{equation}
    {\bf B}({\bf r}) = U_2^{-1}({\bf r})
        \left( \matrix{  \high  {\hat{\bf y}\over 2y^2} & 0 & 0 & 0 \cr\cr
             0 & 0 & 0 & 0 \cr\cr
             0 & 0 & 0 & 0 \cr\cr
             0 & 0 & 0 &  - \high {\hat{\bf y}\over 2y^2}} \right) 
        U_2({\bf r}) +  O(b/y^3) \,\, .
\end{equation}
Thus, at distances large compared to $b$ the non-Abelian part of the
Coulomb magnetic field is cancelled by the cloud in a manner similar
to that which we saw for the SO(5) case.   

In the opposite limit, $b=0$, the solutions are essentially embeddings
of ${\rm SU}(3) \rightarrow {\rm U}(1) \times {\rm U}(1)$ solutions.
At large distances, one finds that
\begin{equation}
    \phi({\bf r}) = U_3^{-1}({\bf r})
        \left( \matrix{ t_4 - \high {1\over 2y_R} & 0 & 0 & 0 \cr\cr
             0 & t_2 - \high{1\over 2y_L} +  \high {1\over 2y_R} & 0 & 0 \cr\cr
             0 & 0 & t_2 & 0 \cr\cr
             0 & 0 & 0 &  t_1 + \high {1\over 2y_L}} \right) 
        U_3({\bf r}) + \cdots
\end{equation}
\begin{equation}
    {\bf B}({\bf r}) = U_3^{-1}({\bf r})
        \left( \matrix{   \high {\bhyr\over 2y_R^2} & 0 & 0 & 0 \cr\cr
             0 & \high {\bhyl\over 2y_L^2} - \high  {\bhyr\over 2y_R^2} & 0
        & 0 \cr\cr  
             0 & 0 & 0 & 0 \cr\cr
             0 & 0 & 0 &   - \high {\bhyl\over 2y_L^2}    } \right) 
        U_3({\bf r}) + \cdots \,\, .
\label{b0asymB}
\end{equation}
Viewed as SU(3) solutions, the long-range fields are purely
Abelian.  Viewed as SU(4) solutions, the long-range part is
non-Abelian in the sense that the unbroken SU(2) acts nontrivially
on the fields.  However, because of the alignment of the fields of the
two massive monopoles, the non-Abelian part of the field is a purely
dipole field that falls as $R/y^3$ at large distances.

\section{Concluding remarks}

I began these lectures by arguing that the one-particle states built
from solitons should not differ in any essential way from those based
on the elementary quanta.  The massless monopoles that arise when
there is nonmaximal symmetry breaking appear to present a challenge to
this point of view.  When viewed in terms of classical solutions, they
do not seem very particle-like: They do not exist as isolated
classical solutions, and in multimonopole configurations they coalesce
into ``clouds'' of arbitrary size rather than appearing as localized
objects with well-defined positions.  Furthermore, while their degrees
of freedom are preserved in the moduli space Lagrangian as one goes
over from the massive to the massless case, the natural interpretation
of these in terms of particle properties are lost.

Of course, the presumed duals to these massless monopoles are the
massless elementary gauge bosons (``gluons'') carrying non-Abelian
charges.  These also differ signficantly from their massive
counterparts, especially in the low-energy regime.  Like the massless
monopoles, these can never be at rest, so it is perhaps not so strange
that there are no are no static solutions corresponding to a single
isolated massless monopole.  Further, the gluon field surrounding a
massive particle carrying non-Abelian charge can be seen as somewhat
analogous to the massless monopole clouds.  It would be very desirable
to be able to make these correspondences more precise, perhaps
including scattering calculations along the lines of those described
in Sec.~4.

Much remains to be learned about the properties and dynamics of these
massless monopoles.  Further investigation of these offers the promise
of deeper insight into the nature of non-Abelian gauge theories.

\vskip .4cm

This work was supported in part by the U.S. Department of Energy.

\end{document}